\documentclass[12pt]{article}

\usepackage{natbib}

\newcommand{\blind}{0}

\usepackage{amssymb}
\usepackage{amsthm}
\usepackage{amsmath}
\usepackage{amsfonts}
\usepackage{booktabs}
\usepackage{graphicx}
\usepackage{tabu}

\usepackage{lineno}
\usepackage[utf8]{inputenc}

\usepackage{booktabs}
\usepackage{here}
\usepackage{dsfont}
\usepackage{float}
\usepackage{amscd}
\usepackage{multirow}
\usepackage{colortbl}
\usepackage[dvipsnames]{xcolor}
\usepackage[margin=1in]{geometry}
\usepackage{setspace}
\usepackage{marginnote}

\usepackage[colorlinks,citecolor=OliveGreen,linkcolor=red,backref,hypertexnames=false]{hyperref}

\graphicspath{{./figures/}}

\newcommand{\eps}{\varepsilon}

\newcommand{\bphi}{\boldsymbol{\phi}}
\newcommand{\bPsi}{\boldsymbol{\Psi}}
\newcommand{\bEta}{\boldsymbol{\eta}}
\newcommand{\bth}{\boldsymbol{\theta}}
\newcommand{\blamb}{\boldsymbol{\lambda}}
\newcommand{\bA}{\mathbf{a}}
\newcommand{\bB}{\mathbf{b}}
\newtheorem{proposition}{Proposition}

\newtheorem{theorem}[proposition]{Theorem}

\theoremstyle{definition}

\theoremstyle{remark}

\IfFileExists{upquote.sty}{\usepackage{upquote}}{}
\begin{document}

\def\spacingset#1{\renewcommand{\baselinestretch}%
{#1}\small\normalsize} \spacingset{1}


\if0\blind
{
  \title{\bf Testing for Threshold Effects in Presence of Heteroskedasticity and Measurement Error with an application to Italian Strikes}
  \author{Francesco Angelini\\
    Department of Statistical Sciences, University of Bologna, Italy\\
    and \\
    Massimiliano Castellani \\
    Department of Statistical Sciences, University of Bologna, Italy\\
    and \\
    Simone Giannerini \\
    Department of Statistical Sciences, University of Bologna, Italy\\
    and \\
    Greta Goracci \\
    Faculty of Economics and Management,\\ Free University of Bozen-Bolzano, Italy
    }
  \maketitle
} \fi

\if1\blind
{
  \bigskip
  \bigskip
  \bigskip
  \begin{center}
    {\LARGE\bf 
    Testing for Threshold Effects in Presence of Heteroskedasticity and Measurement Error with an application to Italian Strikes}
\end{center}
  \medskip
} \fi
\vspace*{-1cm}
\medskip

\begin{abstract}
Many macroeconomic time series are characterised by nonlinearity both in the conditional mean and in the conditional variance and, in practice, it is important to investigate separately these two aspects. Here we address the issue of testing for threshold nonlinearity in the conditional mean, in the presence of conditional heteroskedasticity. We propose a supremum Lagrange Multiplier approach to test a linear ARMA-GARCH model against the alternative of a TARMA-GARCH model. We derive the asymptotic null distribution of the test statistic and this requires novel results since the difficulties of working with nuisance parameters, absent under the null hypothesis, are amplified by the non-linear moving average, combined with GARCH-type innovations. We show that tests that do not account for heteroskedasticity fail to achieve the correct size even for large sample sizes. Moreover, we show that the TARMA specification naturally accounts for the ubiquitous presence of measurement error that affects macroeconomic data. We apply the results to analyse the time series of Italian strikes and we show that the TARMA-GARCH specification is consistent with the relevant macroeconomic theory while capturing the main features of the Italian strikes dynamics, such as asymmetric cycles and regime-switching.
\end{abstract}

\noindent%
\vfill

\newpage
\spacingset{2} 

\section{Introduction} \label{sec:intro}
Strike activity of labour unions is linked to the business cycle and to workers' organizational and political power. Hence, understanding strikes' dynamics can shed light on related social and political issues and support the economic analysis of industrial conflicts in several countries. It is acknowledged that the dynamics of strikes presents complex features, but to the best of our knowledge, both the economic and the econometric literature appears to lack a comprehensive analysis that considers most of these aspects in a unique framework \citep{pal1982,paldam2022,franzosi1989,hundley1987}. The use of strikes as a tool of political and organizational power differs from country to country depending on their specific institutional context \citep{franzosi1989,castellani2013}. In some countries, unions use strikes to influence public policies and the reasons behind them are mostly political, while in other unionised economies, such as Italy, strikes are also used as a bargaining tool in the labour market. Due to its historical and institutional background, the Italian strikes time series is well-suited to investigate the dynamics of this economic and political variable \citep{lange1990,corneo1997,quaranta2012}.
\par
Strike activity has a very long history in Western countries and presents some stylized facts, i.e. they appear cyclically and occur in waves. For example, during the 1970s and 1980s, OECD countries experienced a long strike wave, while a sharp decline in strikes occurred in the last decades \citep{godard2011}. During the 1990s most European countries and the United States witnessed a significant decrease in strike activity, together with a strong shift towards the ``tertiarisation of conflict'' \citep{bordogna2002}. More recently, strikes against governments have been increasingly engaged by unions across Western Europe \citep{hamann2013}. A traditional explanation connects the time series of strikes with the business cycle dynamics. \citet{ree1952} argued that strikes should increase during booms and decrease during the down-swing phase of the business cycle. Wage claims are another cause of strikes.\footnote{\citet{hic1932} pointed out that industrial conflict is costly for both workers and firms. Thus, if agents were rational and fully informed, there would be no reason to strike. Although employees usually strike to gain higher wages or better working conditions, its efficacy remains controversial. \citet{sha1980} surveyed the literature on industrial relations theory.}
However, the empirical evidence for OECD countries seems to indicate that strikes are procyclical and wage increases have a negative effect on strikes.
\par
Qualitative analyses suggest that strikes dynamics is characterized by volatility with clustering effects, primarily due to certain types of strikes that can attract more workers or have prolonged durations \citep{vandaele2016}. Together with the observed asymmetric cyclical behaviour, this hints at the presence of a regime-switching mechanism with conditional heteroskedasticity. We advocate the threshold autoregressive-moving average (TARMA) model with GARCH innovations as a novel and appropriate specification to deal with the aforementioned features of strikes dynamics. Threshold nonlinearity offers a feasible approximation of general complex  dynamics while retaining a good interpretability. Threshold autoregressive models \citep{Ton78,Ton80} have been widely applied in Economics \citep{hansen2011,Cha17b} and Finance \citep{chen2011}. Threshold models are particularly suitable to describe the phenomenon of \textit{regulation} which plays a fundamental role in Finance and Economics. For instance, in financial time series it is common to observe a ``band of inaction'' random walk regime, where arbitrage does not occur, and other regimes where mean reversion takes place so that the model is globally stationary, see e.g. \citet{Cha20b}. Moreover, \cite{pesaran1997} used the TAR model to show that the U.S. GDP is also subject to floor and ceiling effects. \cite{koop1999} estimate TAR models for U.S. unemployment rates and \cite{altissimo2001} propose a VTAR to study the joint dynamics of U.S. GDP and unemployment rates. Note that, none of these studies included a moving-average component in their specifications, most probably due to the lack of developments on non-linear ARMA models. 
\par
TARMA models combine the well-known threshold autoregressive (TAR) model and the threshold moving-average (TMA) model \citep{Lin05}. The incorporation of the moving-average component in a non-linear parametric framework achieves a great approximating capability with few parameters \citep{Gor21}.  They allow to interpret phenomena that change qualitatively across regimes and react differently to shocks, which is a key aspect in macroeconomic dynamics, as also pointed out in \cite{Gon21}. Moreover, as shown in \cite{Cha20b}, they naturally account for the presence of measurement error. Despite these advantages, the theoretical development of TARMA models has been stuck for many years due to unsolved theoretical problems, mainly due to their non-Markovian nature. The impasse has been overcome by \citet{Cha19}, which solved the long-standing open problem regarding the probabilistic structure of the first-order TARMA model. This paved the way for substantial theoretical inferential developments and practical applications.

The main aim of this paper is to introduce a test for threshold ARMA effects with GARCH innovations and ascertain whether the dynamics of the Italian strikes can be adequately modeled using a TARMA-GARCH specification. Existing tests for threshold effects on the conditional mean are negatively affected by the presence of conditional heteroscedasticity. A partial solution to this problem is to use threshold tests sequentially on the residuals of a GARCH model but this is likely to affect non-trivially the overall significance level of the tests. Moreover, the presence of measurement error would require a full TARMA specification but existing tests do not account for this. In particular, \cite{Won97} test an AR-ARCH versus a TAR-ARCH specification, while \cite{Li08} compare an MA-GARCH against a TMA-GARCH. We fill the gap and test the ARMA-GARCH against the TARMA-GARCH specification using a sup-Lagrange Multiplier test tailored to this scope. We derive the asymptotic null distribution of the test statistic and this requires novel results since the inherent difficulties of working with nuisance parameters absent under the null hypothesis are amplified by the non-linear moving average setting combined with GARCH-type innovations. We show that, in presence of heteroskedasticity, tests that assume i.i.d. innovations can be severely biased, whereas our test can be used successfully in all those cases where the aim is testing for a non-linear (threshold) effect in the conditional mean but the series also presents conditional heteroskedasticity.
\par
We use the novel results for the analysis of strikes dynamics, by using Italian data covering the period between January 1949 and December 2009. We identify a regime-switching dynamics with GARCH innovations. We discuss the shortcomings of the linear approach and showcase the adequacy of our non-linear specification, which turns out to be consistent both with observed stylized facts and macroeconomic theories, rooted in the Italian labour market history and on more general social dynamics of labour markets.
\par
The remainder of the paper is organized as follows. In Section~\ref{sec:LM} we introduce our sup-Lagrange multiplier test for threshold effects with GARCH innovations. We present the asymptotic derivations of the null distribution in Section~\ref{sec:null} and, in Section~\ref{sec:sim}, we assess the performance of the test in finite samples by comparing it with the test for threshold nonlinearity that assumes i.i.d. innovations \citep{Gor23}. In Section~\ref{sec:real} we study the dynamics of Italian strikes time series. We apply our test to the monthly series of non-worked hours due to strikes and propose and validate the TARMA-GARCH specification. Finally, Section~\ref{sec:conc} concludes the paper. The Supplementary Material contains all the proofs, additional Monte Carlo results and further analyses of the Italian strikes series.

\section{Test for Threshold Effects with GARCH Innovations}\label{sec:LM}
\subsection{Notation and preliminaries}\label{sec:notation}
Assume the time series $\{X_t:t=0,\pm 1,\pm2,\dots\}$ to follow the threshold autoregressive moving-average model with GARCH errors,  TARMA$(p,q)$-GARCH$(u,v)$, defined by the system of difference equations:
\begin{align}\label{eq:TARMA-GARCH}
  X_t &=\phi_{0} + \sum_{i=1}^{p}\phi_{i}X_{t-i} - \sum_{j=1}^{q}\theta_{j} \eps_{t-j} + \eps_t\nonumber\\
      &+\left(\varphi_{0} + \sum_{i=1}^{p}\varphi_{i}X_{t-i} - \sum_{j=1}^{q}\vartheta_{j} \eps_{t-j}\right)I(X_{t-d}\leq r)\\
\eps_t&=\sqrt{h_t}z_t; \quad
h_t   =a_0 + \sum_{i=1}^{u}a_i\eps^2_{t-i} + \sum_{j=1}^{v}b_j h_{t-j}.\nonumber
\end{align}
\noindent
The process $\{z_t\}$ is a sequence of i.i.d. random variables with zero mean, unit variance and finite fourth moment. $p$ and $q$ are the autoregressive and moving-average orders, respectively; $d$ is the delay parameter; $u$ and $v$ are the ARCH and GARCH orders, respectively. We assume $p,q,d,u,v$ to be known positive integers. Moreover, $I(\cdot)$ is the indicator function and $r\in\mathds{R}$ is the threshold parameter. For notational convenience, we abbreviate $I(X_t\leq r)$ by $I_r(X_t)$. We define the following vectors containing the parameters of Model~(\ref{eq:TARMA-GARCH}):
$\bphi = \left(\phi_{0},\phi_{1},\dots,\phi_{p}\right)^\intercal\in\Theta_\phi,$ $\boldsymbol\varphi=\left(\varphi_{0},\varphi_{1},\dots,\varphi_{p}\right)^\intercal\in\Theta_\varphi,$  
  $\bth = \left(\theta_{1},\dots,\theta_{q}\right)^\intercal\in\Theta_\theta,$ $\boldsymbol\vartheta=\left(\vartheta_{1},\dots,\vartheta_{q}\right)^\intercal\in\Theta_\vartheta,$ 
  $\bA = \left(a_0,a_1,\dots,a_u\right)^\intercal\in\Theta_a,$  $\bB = \left(b_1,\dots,b_v\right)^\intercal\in\Theta_b,$
with $\Theta_\phi,\Theta_\varphi\subseteq\mathds{R}^{p+1}$; $\Theta_\theta,\Theta_\vartheta\subseteq\mathds{R}^{q}$; $\Theta_a\subseteq\mathds{R}^{u+1}$ and $\Theta_b\subseteq\mathds{R}^{v}$. Moreover let
\begin{align}\label{eq:pars}
&\bPsi_1 = (\bphi^\intercal,\bth^\intercal)^\intercal,\quad \bPsi_2 = (\boldsymbol\varphi^\intercal,\boldsymbol\vartheta^\intercal)^\intercal,
\quad \bPsi=(\bPsi_1^\intercal,\bPsi_2^\intercal)^\intercal,\quad \blamb=\left(\bPsi_1^\intercal,\bA^\intercal,\bB^\intercal\right)^\intercal,\quad \boldsymbol\eta=\left(\blamb^\intercal,\bPsi_2^\intercal\right)^\intercal.
\end{align}
$\bPsi_2$ and $\bPsi_1$ contain the ARMA parameters to be tested and those who are not, respectively; $\blamb$ is the parametric vector for the ARMA-GARCH model whereas $\boldsymbol\eta$ is the vector of all the parameters (excluding the threshold $r$) in Model~(\ref{eq:TARMA-GARCH}).
We use $\bPsi$, $\bphi$, etc.$\dots$ to refer to unknown parameters, whereas the true parameters are obtained by adding the $0$ subscript, i.e.:
\noindent
$\bPsi_0=(\bPsi_{0,1}^\intercal,\bPsi_{0,2}^\intercal)^\intercal$, $\blamb_0^\intercal$ and $\boldsymbol\eta_0^\intercal$. Also, we assume $\boldsymbol\eta_0$ to be an interior point of the parameter space.
\par
 We test whether a TARMA$(p,q)$-GARCH$(u,v)$ model provides a significantly better fit than the linear ARMA$(p,q)$-GARCH$(u,v)$ model by developing a Lagrange multiplier test statistics. Letting $\boldsymbol 0$ be the vector of all zeroes, the system of hypothesis results:
\begin{equation}\label{eq:H}
\begin{cases}
  H_0&:\bPsi_{0,2}=\boldsymbol 0 \\
  H_1&: \bPsi_{0,2}\neq\boldsymbol 0,
\end{cases}
\end{equation}
Under $H_0$ the process follows a linear ARMA$(p,q)$-GARCH$(u,v)$ model:
\begin{align}\label{eq:ARMA-GARCH}
  X_t &=\phi_{0,0} + \sum_{i=1}^{p}\phi_{0,i}X_{t-i} - \sum_{j=1}^{q}\theta_{0,j} \eps_{t-j} + \eps_t\\
\eps_t&=\sqrt{h_t}z_t; \quad
h_t   =a_{0,0} + \sum_{i=1}^{u}a_{0,i}\eps^2_{t-i} + \sum_{j=1}^{v}b_{0,j} h_{t-j}.\nonumber
\end{align}
Define the polynomials
$\phi(z)=1-\phi_{1}z-\phi_{2}z^2-\dots-\phi_{p}z^p$, $\theta(z)=1-\theta_{1}z-\theta_{2}z^2-\dots-\theta_{q}z^q,$
$a(z)=1-a_{1}z-a_{2}z^2-\dots-a_{u}z^u$, $b(z)=1-b_{1}z-b_{2}z^2-\dots-b_{v}z^v,$
$\varphi(z)=1-\varphi_{1}z-\varphi_{2}z^2-\dots-\varphi_{p}z^p,$ $\vartheta(z)=1-\vartheta_{1}z-\vartheta_{2}z^2-\dots-\vartheta_{q}z^q.$
We assume the following:
\begin{description}
  \item[Assumptions] \phantom{bla}
   \begin{itemize}
     \item [A.1] $\phi(z)\neq0$ and $\theta(z)\neq 0$ for all $z\in\mathds{C}$ such that $|z|\leq 1$ and they do not share common roots. $\varphi(z)$ and $\vartheta(z)$ are also coprime.
     \item [A.2] $a_i>0$, $i=0,1,\dots,u$; $b_j>0$, $j=1,\dots,v$; $|\sum_{i=1}^{u}a_i+\sum_{j=1}^{v}b_j|<1$; $a(z)$ and $b(z)$ are coprime.
     \item [A.3] $\{z_t\}$ is a sequence of i.i.d. random variables with $E[z_t]=0$, $E[z_t^2]=1$ and $E[z_t^4]<\infty$. Moreover, $z_t$ has a continuous and positive density function, say $f_z(x)$.
     \item [A.4] The sequence $\{\eps_t\}$ is strictly stationary and ergodic with a finite fourth moments.
     \item [A.5] The process $\{X_t\}$ is ergodic and invertible under $H_0$.
   \end{itemize}
\end{description}
These assumptions are common in deriving the asymptotic behaviour of test for threshold nonlinearity. See, inter alia, \citet{Li08}, \citet{Gor23}, \citet{Cha90a}. In particular, Assumptions A.1 and A.2 allow to estimate and identify uniquely the parameter of the ARMA and GARCH part, respectively. These assumptions also imply that the process $\{h_t\}$ is strictly stationary and ergodic with $E[h_t^2]<\infty$ and it is bounded away from zero with probability 1, see \citet{Li08} and \citet{Li11} for further details.
\par
Suppose we observe $X_1,\dots,X_n$. Omitting a negative constant, the Gaussian log-likelihood conditional on the initial values $X_0,X_{-1},\dots$ is:

\begin{align}
\ell_n(\bEta,r)&=-\frac{1}{2}\sum_{t=1}^n \frac{\eps_t^2(\bEta,r)}{h_t(\bEta,r)}+\ln(h_t(\bEta,r)),\label{eq:ell}\\
\text{where }\; \eps_t(\bEta,r)&=X_t - \left\{\phi_{0} + \sum_{i=1}^{p}\phi_{i}X_{t-i} - \sum_{j=1}^{q}\theta_{j}\eps_{t-j}(\bEta,r)\right\} \nonumber\\
&-\left\{\varphi_{0} + \sum_{i=1}^{p}\varphi_{i}X_{t-i} - \sum_{j=1}^{q}\vartheta_{j}\eps_{t-j}(\bEta,r)\right\} I_r\left(X_{t-d}\right)\label{eq:eps}\\
h_t(\bEta,r)&= a_0 + \sum_{i=1}^{u}a_i\eps^2_{t-i}(\bEta,r) + \sum_{j=1}^{v}b_j h_{t-j}(\bEta,r).\label{eq:h}
\end{align}
\noindent
Also:
\begin{align}
  \eps_t(\blamb)&= \eps_t(\bEta,-\infty) = X_t-\left\{\phi_{0}+ \sum_{i=1}^{p}\phi_{i}X_{t-i}- \sum_{j=1}^{q}\theta_{j}\eps_{t-j}(\blamb)\right\};\label{eq:eps_0}\\
  h_t(\blamb)&=h_t(\bEta,-\infty)= a_0 + \sum_{i=1}^{u}a_i\eps^2_{t-i}(\blamb) + \sum_{j=1}^{v}b_j h_{t-j}(\blamb),\label{eq:h_0}
\end{align}
\noindent
and, under the null hypothesis, $\eps_t(\bEta_0,r)=\eps_t$ and $h_t(\bEta_0,r)=h_t$.
\par

The derivation of the Lagrange multipliers test is based upon the first and second partial derivatives $\ell_n(\bEta,r)$ with respect to $\bPsi$. Let
\begin{align*}
&\frac{\partial\ell_n(\bEta,r)}{\partial\bPsi}= \left(\left(\frac{\partial\ell_n(\bEta,r)}{\partial\bPsi_1}\right)^\intercal, \left(\frac{\partial\ell_n(\bEta,r)}{\partial\bPsi_2}\right)^\intercal\right)^\intercal\\
&=\sum_{t=1}^{n}\left\{- \frac{\eps_t(\bEta,r)}{h_t(\bEta,r)}\frac{\partial \eps_t(\bEta,r) }{\partial\bPsi} + \frac{1}{2}\left(\frac{\eps_t^2(\bEta,r)}{h_t^2(\bEta,r)} - \frac{1}{h_t(\bEta,r)}\right) \frac{\partial h_t(\bEta,r) }{\partial\bPsi}  \right\}
\end{align*}
\noindent
with $\partial\eps_t(\bPsi,r)/\partial\bPsi$ (respectively $\partial h_t(\bPsi,r)/\partial\bPsi$) being the partial derivative of $\eps_t(\bPsi,r)$  ($h_t(\bPsi,r)$) with respect to $\bPsi$:
\begin{align*}
&\frac{\partial \eps_t(\bEta,r) }{\partial\bPsi}=D_t+\sum_{j=1}^{q}\theta_j\frac{\partial \eps_{t-j}(\bEta,r) }{\partial\bPsi},\\
\text{with }D_t=&\left(-1,-X_{t-1},\dots,-X_{t-p},\eps_{t-1},\dots,\eps_{t-q},\right.\\
&\left.-I_r(X_{t-d}),-X_{t-1}I_r(X_{t-d}),\dots,-X_{t-p}I_r(X_{t-d}),\eps_{t-1}I_r(X_{t-d}),\dots,\eps_{t-q}I_r(X_{t-d})\right)^\intercal\\
&\frac{\partial h_t(\bEta,r) }{\partial\bPsi} = 2\sum_{i=1}^{u}a_i\eps_{t-i}\frac{\partial \eps_{t-i}(\bEta,r) }{\partial\bPsi} + \sum_{j=1}^{v}b_j\frac{\partial h_{t-i}(\bEta,r) }{\partial\bPsi}.
\end{align*}
\noindent
 Lastly, define the block matrix $\mathcal{I}_n(\bEta,r)$:
\begin{equation}\label{eq:Fmatrix}
\mathcal{I}_n(\bEta,r)
      =\begin{pmatrix}
                  \mathcal{I}_{n,11}(\bEta)    & \mathcal{I}_{n,12}(\bEta,r) \\
                  \mathcal{I}_{n,21}(\bEta,r) & \mathcal{I}_{n,22}(\bEta,r)
       \end{pmatrix}
  =\begin{pmatrix}
              -\frac{\partial^2\ell_n(\bEta,r)}{\partial\bPsi_1\partial\boldsymbol\bPsi_1^\intercal}   & -\frac{\partial^2\ell_n(\bEta,r)}{\partial\bPsi_1\partial\boldsymbol\bPsi_2^\intercal} \\
              -\frac{\partial^2\ell_n(\bEta,r)}{\partial\bPsi_2\partial\boldsymbol\bPsi_1^\intercal} &
              -\frac{\partial^2\ell_n(\bEta,r)}{\partial\bPsi_2\partial\boldsymbol\bPsi_2^\intercal}
   \end{pmatrix}.
\end{equation}
The Lagrange multiplier approach requires estimating the model under the null hypothesis. Hence, let
$\hat{\blamb}=(\hat{\bphi}^\intercal, \hat{\bth}^\intercal, \hat{\bA}^\intercal, \hat{\bB}^\intercal)^\intercal=\arg\min_{\blamb}\ell_n(\blamb)$, 
with $\ell_n(\blamb)=\ell_n(\bEta,-\infty)$.
 Hence, $\hat{\blamb}$ is the Maximum Likelihood Estimator (hereafter MLE) of the ARMA-GARCH coefficients in Eq.~(\ref{eq:ARMA-GARCH}) and we define $\hat{\bEta}=(\hat{\blamb}^\intercal,\boldsymbol 0^\intercal)^\intercal$
  to be the so called \textit{restricted} MLE, i.e. under the null hypothesis. We write $\partial \hat{\ell}_n(r)/\partial \bPsi_2$ and $\hat{\mathcal{I}}_{n}(r)$ to refer to  $\partial \ell_n(\bEta,r)/\partial \bPsi$ and  $\mathcal{I}_n(\bEta,r)$ evaluated at the restricted MLE $\hat{\bEta}$ , i.e.:
\begin{align*}
\frac{\partial \hat{\ell}_n(r) }{\partial \bPsi_2} =\frac{\partial\ell_n(\hat{\bEta},r)}{\partial\bPsi_2}; \qquad
\hat{\mathcal{I}}_n(r) = \mathcal{I}_n(\hat{\bEta},r)
      =\begin{pmatrix}
                  \hat{\mathcal{I}}_{n,11}    & \hat{\mathcal{I}}_{n,12}(r) \\
                  \hat{\mathcal{I}}_{n,21}(r) & \hat{\mathcal{I}}_{n,22}(r)
       \end{pmatrix}.
\end{align*}
Under the null hypothesis, the threshold parameter $r$ is absent thereby the standard asymptotic theory is not applicable. To cope with this issue, we firstly develop the Lagrange multiplier test statistic as a function of $r$ ranging in a data-driven set $\mathcal{R}=[r_L,r_U]$, with $r_L$ and $r_U$ being, e.g., some percentiles of the data. Then, we take the overall test statistic as the supremum on $\mathcal{R}$. This approach has become widely used in the literature of tests involving nuisance parameters. This was first proposed in the seminal work of \citet{Cha90a}, and followed by \cite{And93} and \citet{Han96}. Within the threshold setting, the idea was deployed in \citet{Won97}, \citet{Lin05}, \citet{Li08}, \citet{Li11}, \citet{Cha20b}, \citet{Gor21}. Recently, \citet{Gia23} adapted it to prove the validity of a bootstrap scheme in testing threshold nonlinearity.
\par\noindent
The test statistic is
  \begin{align}
  T_n&=\sup_{r\in[r_L,r_U]}T_{n}(r), \label{eq:Tn}\\
 T_{n}(r)&=\left(\frac{\partial \hat{\ell}_n(r) }{\partial \boldsymbol\Psi_2}\right)^\intercal \left(\hat{\mathcal{I}}_{n,22}(r)-\hat{\mathcal{I}}_{n,21}(r)\hat{\mathcal{I}}_{n,11}^{-1} \hat{\mathcal{I}}_{n,12}(r)\right)^{-1}\frac{\partial \hat{\ell}_n(r) }{\partial \boldsymbol\Psi_2}.\label{eq:Tnr}
  \end{align}
\noindent
In Eq.~(\ref{eq:Tn}), besides taking the supremum, other convenient functions can be used to derive an overall test statistic.
\subsection{The Null Distribution}\label{sec:null}
In this section we derive the asymptotic distribution of $T_n$ under the null hypothesis that $\{X_t\}$ follows the ARMA$(p,q)$-GARCH$(u,v)$ process defined in Eq.~(\ref{eq:ARMA-GARCH}). Hereafter, all the expectations are taken under the true probability distribution for which $H_0$ holds. We use  $\|\cdot\|$ to refer to the $\mathcal{L}^2$ matrix norm (the Frobenius' norm, i.e. $\|A\|=\sqrt{\sum_{i=1}^{n}\sum_{j=1}^{m}|a_{ij}|^2}$, where $A$ is a $n\times m$ matrix). Also, $o_p(1)$ indicates the convergence in probability to zero as $n$ increases.  $\mathcal{D}_\mathds{R}(a,b)$, $a<b$, is the space of functions from $(a,b)$ to $\mathds{R}$ that are right continuous with left-hand limits. We assume  $\mathcal{D}_\mathds{R}(a,b)$ to be equipped with the topology of uniform convergence on compact sets, see \citet{Bil68} for further  details.
\par\noindent
 In order to obtain its asymptotic distribution in the main theorem, we derive, under the null hypothesis, an (asymptotic) \textit{uniform} approximation of the test statistic depending on the true parameters $\bEta_0=(\blamb_0^\intercal,\boldsymbol 0^\intercal)^\intercal$. In this respect, define the vector $ \nabla_n(r)=\left(\nabla^\intercal_{n,1},\nabla^\intercal_{n,2}(r)\right)^\intercal$, with
 \begin{align*}
 \nabla_{n,1}&=\frac{1}{\sqrt{n}}\sum_{t=1}^{n}\left\{- \frac{\eps_t}{h_t}\frac{\partial \eps_t(\bEta_0,r) }{\partial\bPsi_1} + \frac{1}{2}\left(\frac{\eps_t^2}{h_t^2} - \frac{1}{h_t}\right) \frac{\partial h_t(\bEta_0,r) }{\partial\bPsi_1}  \right\},\\
 \nabla_{n,2}(r)&=\frac{1}{\sqrt{n}}\sum_{t=1}^{n}\left\{- \frac{\eps_t}{h_t}\frac{\partial \eps_t(\bEta_0,r) }{\partial\bPsi_2} + \frac{1}{2}\left(\frac{\eps_t^2}{h_t^2} - \frac{1}{h_t}\right) \frac{\partial h_t(\bEta_0,r) }{\partial\bPsi_2}  \right\}
 \end{align*}
 and the matrix
 \begin{align*}
\Lambda(r)=\begin{pmatrix}
\Lambda_{11} & \Lambda_{12}(r) \\
\Lambda_{21}(r) & \Lambda_{22}(r)
\end{pmatrix}
= E\left[\frac{1}{h_t} \left(\frac{\partial \eps_t(\bEta_0,r) }{\partial\bPsi}\right) \left(\frac{\partial \eps_t(\bEta_0,r) }{\partial\bPsi}\right)^\intercal+ \frac{1}{2h_t} \left(\frac{\partial h_t(\bEta_0,r) }{\partial\bPsi}\right) \left(\frac{\partial h_t(\bEta_0,r) }{\partial\bPsi}\right)^\intercal\right].
\end{align*}
In the following proposition, under the null hypothesis and Assumption A, we provide a uniform approximation that will allow us to derive the asymptotic null distribution of the test statistic $T_n$ in the main theorem.
\begin{proposition}\label{prop:asymptotics}
Under Assumptions A.1--A.5 and under $H_0$, we have the following:
\begin{description}
  \item[(i)]
  \begin{align*}
  \sup_{r\in[a,b]}  \left\|\left(\frac{\hat{\mathcal{I}}_{n,22}(r)}{n} - \frac{\hat{\mathcal{I}}_{n,21}(r)}{n} \left(\frac{\hat{\mathcal{I}}_{n,11}}{n}\right)^{-1}
  \frac{\hat{\mathcal{I}}_{n,12}(r)}{n}\right)^{-1}
  -\left(\Lambda_{22}(r)- \Lambda_{21}(r)\Lambda_{11}^{-1}\Lambda_{12}(r)\right)^{-1}\right\| = o_p(1).
  \end{align*}
  \item[(ii)] $$\sup_{r\in[a,b]}\left\|\frac{1}{\sqrt{n}}\frac{\partial\hat\ell_n(r)}{ \partial\boldsymbol\Psi_2}- \left(\nabla_{n,2}(r)-\Lambda_{21}(r)\Lambda_{11}^{-1}\nabla_{n,1}\right)\right\| =o_p(1).$$
\end{description}
\end{proposition}
\noindent
Define the process:
$\{Q(r),r\in\mathds{R}\}$, with $Q(r)=\left(\nabla_{n,2}(r)-\Lambda_{21}(r)\Lambda_{11}^{-1}\nabla_{n,1}\right)$.
Note that $\{Q(r)\}$ is a  marked empirical process with infinitely many markers. In the next theorem we derive a novel Functional Central Limit Theorem (hereafter FCLT) for $\{Q(r)\}$.
\begin{theorem}\label{th:FCLT}
 Let  $\left\{\xi(r),\;r\in\mathds{R}\right\}$ be a centered Gaussian vector process of dimension  $(p+q+1)$, with covariance kernel
 $\Sigma(r,s) = \Lambda_{22}(r\wedge s)-\Lambda_{21}(r)\Lambda_{11}^{-1}\Lambda_{12}(s).$
 Under Assumptions  A.1--A.5 and $H_0$, $Q(r)$ converges weakly to $\xi(r)$ in $D_{p+q+1}(-\infty,+\infty)$.
\end{theorem}
\noindent
By invoking the continuous mapping theorem, the asymptotic null distribution of our Lagrange multiplier test statistic readily follows.
 \begin{theorem}\label{th:null}
    Under Assumptions  A.1--A.5 and $H_0$, asymptotically, the Lagrange multiplier test statistic $T_n$ has the same distribution of
   \begin{equation}\label{eqn:sup}
     \sup_{r\in[r_L,r_U]}\xi(r)^\intercal\Sigma(r,r)^{-1}\xi(r),
   \end{equation}
   where $\xi(r)$ and  $\Sigma(r,s)$ are defined in Theorem~\ref{th:FCLT}.
 \end{theorem}

\section{Finite Sample Performance}\label{sec:sim}

In this section, we study the finite sample performance of our test, which we denote by sLMg, and compare it with the sLM test of \cite{Gor23}, which assumes i.i.d. innovations. The length of the series is $n=100,200,500$ and $z_t$, $t=1,\dots,n$ is generated from a standard Gaussian white noise. The nominal size of the tests is $\alpha=5\%$ and the number of Monte Carlo replications is 10000. Furthermore, we use the tabulated critical values of \cite{And03} and the threshold is searched from percentile 25th to 75th of the sample distribution.

\subsection{Size}\label{sec:size}
\begin{figure}
  \centering
\includegraphics[width=0.8\linewidth,keepaspectratio]{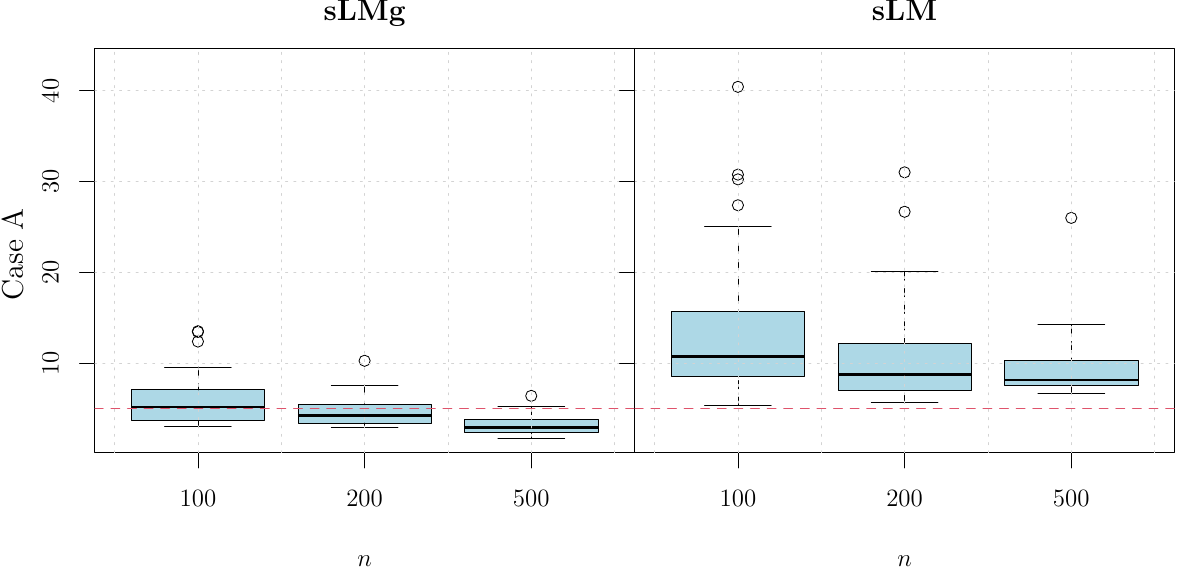}
\includegraphics[width=0.8\linewidth,keepaspectratio]{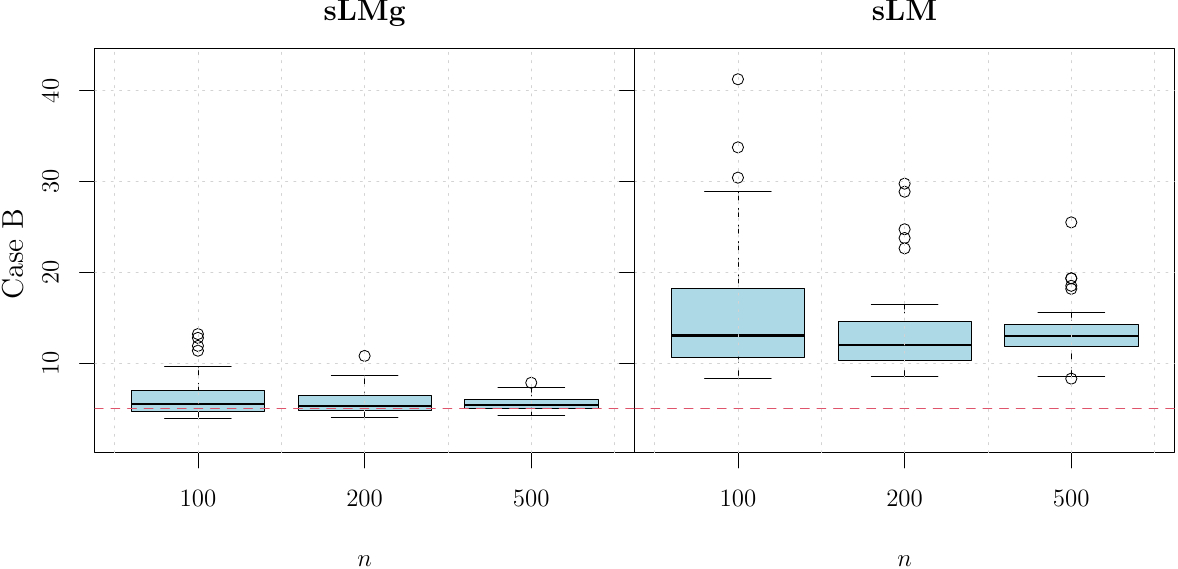}
\includegraphics[width=0.8\linewidth,keepaspectratio]{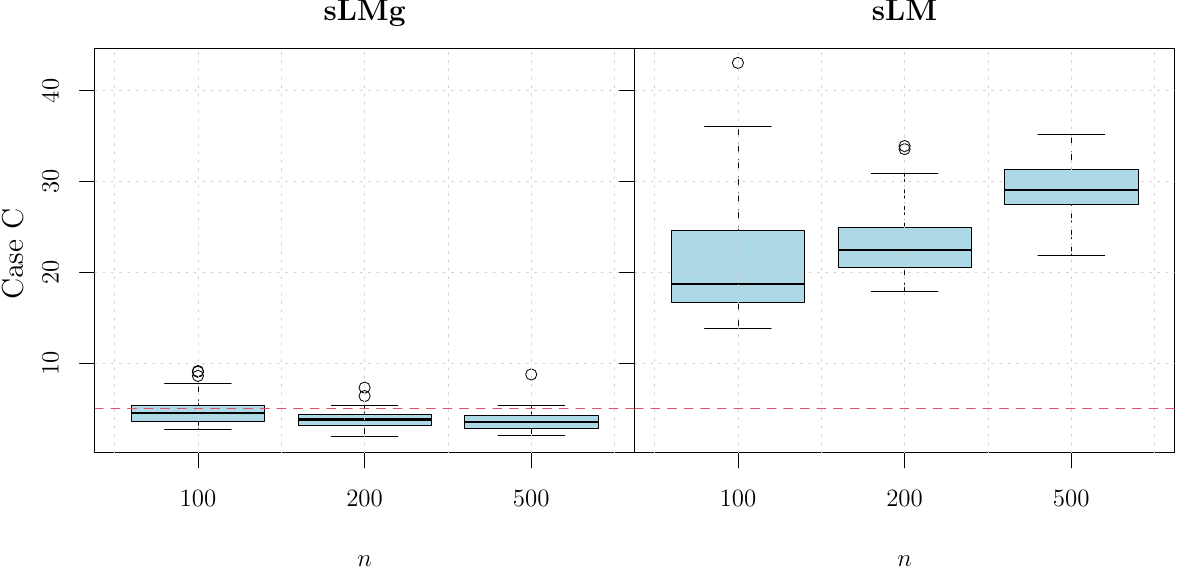}
 \caption{Empirical size (percent) of the sLMg and sLM tests, at nominal level $\alpha=5\%$ for the ARMA(1,1)-GARCH(1,1) process of Eq.~(\ref{ARMAGARCHsim}). Cases A,B,C.}\label{fig:size1}
\end{figure}

We study the empirical size by simulating from the following ARMA$(1,1)$-GARCH$(1,1)$ model:
\begin{align}\label{ARMAGARCHsim}
 X_t &= \phi_1 X_{t-1} + \theta_1 \eps_{t-1} + \eps_t.\nonumber\\
 \eps_t&=\sqrt{h_t}z_t; \quad h_t=a_0 + a_1\eps^2_{t-1} + b_1 h_{t-1}.
\end{align}
where $\phi_1=(-0.9,-0.6,-0.3,0.0,0.3,0.6,0.9)$ and $\theta_1=(-0.8,-0.4,0.0,0.4,0.8)$. We combine these with the following parameters for the GARCH specification:  $(a_0,a_1,b_1) = (1,0.04,0.95)$ (case A), $(1,0.3,0.0)$ (case B), $(1,0.4,0.4)$ (case C), so as to obtain 35 different parameter settings for each case.  Notice that case B corresponds to an ARCH(1) process. Also, all three cases fulfil the condition of finite fourth moments. The results are presented in Figure~\ref{fig:size1}, where the boxplots group together the 35 configurations for each of the three cases. The full results are reported in Tables~\ref{tab:A}--\ref{tab:C} of the Supplementary Material. Clearly, the standard sLM test is oversized and, especially for case C, the bias increases with the sample size and can be severe. On the contrary, our TARMA-GARCH test has correct size in almost every setting for a sample size as low as  $n=100$. Note that this holds also for corner cases, such as near integrated cases or when near cancellation of the AR and MA polynomials occurs. As the sample size increases, the empirical size tends to concentrate around the nominal $5\%$ for case B, whereas it seems to settle around lower values for cases A and C. As we will show in the next section, this slight undersize does not impinge negatively upon the power and results in a conservative test, which is generally appealing in practical applications. The size for the three cases grouped together is reported in Figure~\ref{fig:size2} of the Supplementary Material.

\subsection{Power}\label{sec:power}
\begin{figure}
  \centering
\includegraphics[width=0.45\linewidth,keepaspectratio]{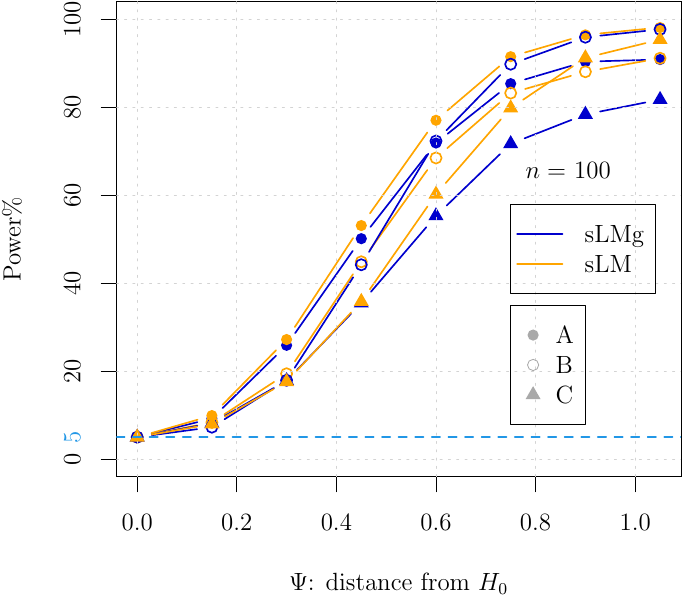}
\includegraphics[width=0.45\linewidth,keepaspectratio]{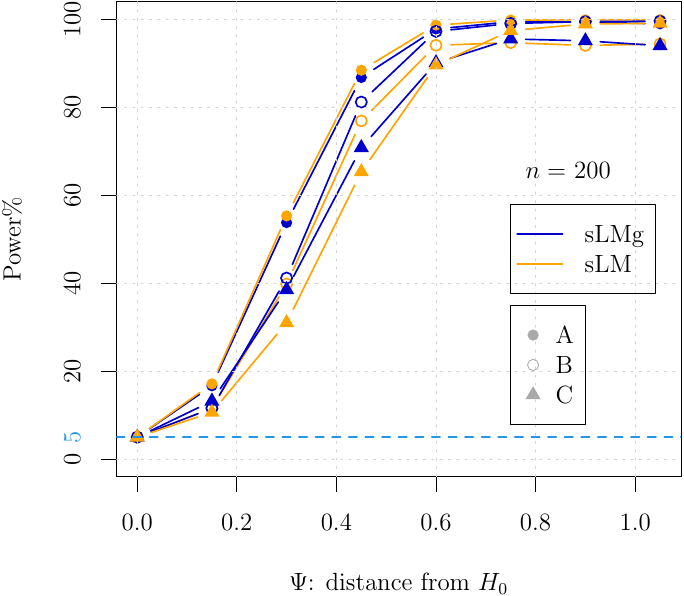}
\includegraphics[width=0.45\linewidth,keepaspectratio]{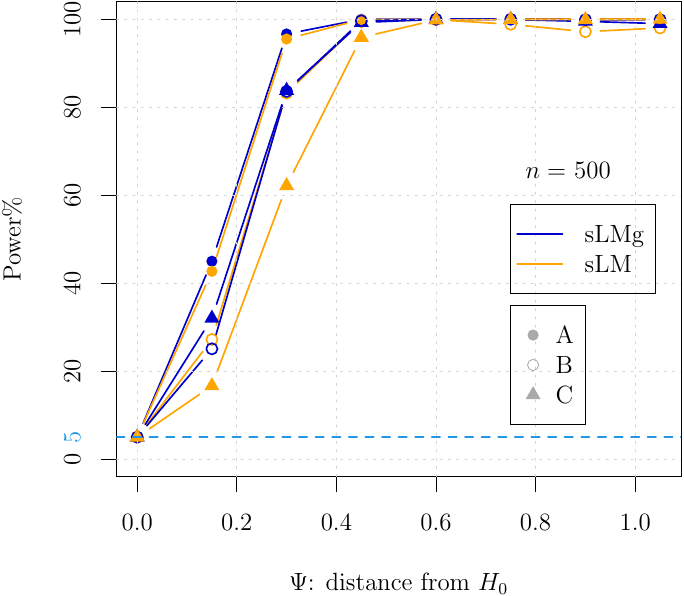}
  \caption{Size corrected power (percent) of the sLMg and sLM tests, at nominal level $\alpha=5\%$ for the TARMA(1,1)-GARCH(1,1)  process of Eq.~(\ref{TARMAGARCHsim}).}\label{fig:3b}
\end{figure}

In order to study the power of the tests we simulate from the following TARMA$(1,1)$-GARCH$(1,1)$ model:
\begin{align}\label{TARMAGARCHsim}
 X_t &= 0.5 + 0.5 X_{t-1} + 0.5 \eps_{t-1} + \left(\varphi_0 + \varphi_1 X_{t-1} + \vartheta_1\eps_{t-1}\right)I(X_{t-1}\leq 0) + \eps_t.\nonumber\\
 \eps_t&=\sqrt{h_t}z_t; \quad h_t=a_0 + a_1\eps^2_{t-1} + b_1 h_{t-1}.
\end{align}
where $\varphi_{0}=\varphi_1=\vartheta_1 = 0.5 + \Psi$, where $\Psi = (0.00, -0.15, -0.30, -0.45, -0.60, -0.75, -0.90, -1.05).$
Hence, the system of hypotheses of Eq.~(\ref{eq:H}) becomes

$$
\begin{cases}
  H_0&:\bPsi= \mathbf{0} \\
  H_1&:\bPsi\neq\mathbf{0},
\end{cases}
$$
\noindent
where $\bPsi = (\Psi,\Psi,\Psi)$, so that the parameter $\Psi$  represents the departure from the null hypothesis. As above, we combine these with the following parameters for the GARCH specification:  $(a_0,a_1,b_1) = (1,0.1,0.8)$ (case A), $(1,0.4,0.4)$ (case B), $(1,0.8,0.1)$ (case C). The size-corrected power of the tests (in percentage) is presented in Figure~\ref{fig:3b}, where each panel corresponds to a different sample size. The blue and orange lines correspond to the sLMg and sLM test, respectively, whereas the different points correspond to cases A (filled round dot), B (empty round dot) and C (filled triangle). The behaviour of the two tests is very similar, and the power loss incurred by estimating the GARCH parameters in the sLMg test is very small and is overwhelmingly compensated by the correct size in presence of heteroskedasticity. The full results are reported in the Supplementary Material, Table~\ref{tab:4} (size corrected power) and Table~\ref{tab:4u} (raw power). The general conclusion that can be drawn is that, even in absence of information regarding the presence of heteroskedasticity, it is safe to use the sLMg test as it will lower the risk of a false rejection while retaining a good discriminating power.

\subsection{Measurement Error}\label{SMsec:Merr}
We assess the effect of measurement error on the size of the tests. We simulate $X_t$ from the following AR(1)-GARCH(1,1) model
\begin{align}\label{ARGARCHsim}
 X_t &= \phi_1 X_{t-1} +  \eps_t.\nonumber\\
 \eps_t&=\sqrt{h_t}z_t; \quad
h_t=a_0 + a_1\eps^2_{t-1} + b_1 h_{t-1},
\end{align}
where, as above, $\phi_1=(-0.9,-0.6,-0.3,0.0,0.3,0.6,0.9)$ and  $(a_0,a_1,b_1) = (1,0.04,0.95)$ (case A), $(1,0.3,0.0)$ (case B), $(1,0.4,0.4)$ (case C). We add measurement noise as follows: $Y_t = X_t + \eta_t$, 
where the measurement error $\eta_t\sim N(0,\sigma^2_\eta)$ is such that the signal to noise ratio SNR~$=\sigma^2_X/\sigma^2_\eta$ is equal to $\{\infty, 50,10,5\}$. Here, $\sigma^2_X$ is the unconditional variance of $X_t$ computed by means of simulation. The case without noise (SNR $=\infty$) is taken as the benchmark. The empirical size (rejection percentages) for the three sample sizes is presented in Table~\ref{tab:MEtot}. Clearly, the size of the sLMg test is either minimally affected or not affected at all by the presence of measurement error, even for high levels of noise (signal to noise ratio = 5).
\begin{table}
\centering
\caption{\label{tab:MEtot}Empirical size of the sLMg and sLM tests at nominal level $\alpha=5\%$ for the AR(1)-GARCH(1,1)  process of Eq.~(\ref{ARGARCHsim}) with three levels of measurement error.}
\begin{tabular}[t]{lrrrrrrrrrrrrr}
\multicolumn{2}{c}{ } & \multicolumn{4}{c}{$n=100$} & \multicolumn{4}{c}{$n=200$} & \multicolumn{4}{c}{$n=500$} \\
\cmidrule(l{3pt}r{3pt}){3-6} \cmidrule(l{3pt}r{3pt}){7-10} \cmidrule(l{3pt}r{3pt}){11-14}
  & $\phi_1$ & $\infty$ & 50 & 10 & 5 & $\infty$ & 50 & 10 & 5 & $\infty$ & 50 & 10 & 5\\
\cmidrule(l{3pt}r{3pt}){3-6} \cmidrule(l{3pt}r{3pt}){7-10} \cmidrule(l{3pt}r{3pt}){11-14}
 & -0.9 & 3.7 & 2.6 & 2.8 & 4.2 & 3.2 & 2.7 & 3.8 & 4.5 & 2.7 & 2.8 & 4.6 & 5.1\\
 & -0.6 & 3.5 & 4.0 & 3.7 & 4.5 & 3.0 & 3.2 & 3.6 & 3.7 & 2.5 & 2.1 & 2.1 & 3.1\\
 & -0.3 & 4.7 & 5.4 & 5.4 & 6.8 & 3.0 & 2.3 & 3.0 & 4.1 & 2.0 & 2.7 & 2.1 & 3.1\\
A & 0.0 & 7.7 & 8.4 & 9.3 & 8.7 & 7.5 & 7.9 & 7.8 & 8.0 & 4.2 & 2.4 & 3.8 & 4.1\\
 & 0.3 & 4.0 & 4.8 & 3.8 & 4.6 & 2.8 & 2.5 & 3.1 & 3.4 & 1.3 & 1.4 & 1.7 & 2.2\\
 & 0.6 & 3.0 & 2.8 & 3.4 & 3.4 & 2.9 & 3.7 & 3.2 & 3.4 & 2.0 & 2.8 & 2.5 & 3.6\\
 & 0.9 & 3.3 & 3.7 & 4.2 & 5.0 & 3.4 & 4.1 & 4.3 & 4.7 & 2.8 & 2.3 & 2.6 & 3.5\\
\addlinespace
 & -0.9 & 3.7 & 3.3 & 3.0 & 3.6 & 4.8 & 4.3 & 4.3 & 4.9 & 3.5 & 5.3 & 4.9 & 4.8\\
 & -0.6 & 4.4 & 3.9 & 4.1 & 3.9 & 4.5 & 4.5 & 6.1 & 5.3 & 6.6 & 5.9 & 6.2 & 6.8\\
 & -0.3 & 4.8 & 5.6 & 7.0 & 5.0 & 4.5 & 5.5 & 6.3 & 5.7 & 5.5 & 4.1 & 5.7 & 6.3\\
B & 0.0 & 9.4 & 9.6 & 7.8 & 9.8 & 6.4 & 6.4 & 7.1 & 7.1 & 7.6 & 8.1 & 7.2 & 9.2\\
 & 0.3 & 4.9 & 4.2 & 5.0 & 5.1 & 3.1 & 4.1 & 5.5 & 5.2 & 5.3 & 4.9 & 4.3 & 6.1\\
 & 0.6 & 4.0 & 4.4 & 4.1 & 4.4 & 4.4 & 4.3 & 4.7 & 5.0 & 6.3 & 5.3 & 5.6 & 6.5\\
 & 0.9 & 4.3 & 4.8 & 4.6 & 5.1 & 4.5 & 4.3 & 3.9 & 3.5 & 5.2 & 5.4 & 5.5 & 5.8\\
\addlinespace
 & -0.9 & 3.3 & 2.9 & 3.8 & 5.7 & 2.5 & 3.6 & 4.6 & 4.1 & 3.7 & 5.6 & 5.9 & 6.3\\
 & -0.6 & 2.4 & 2.1 & 3.3 & 4.0 & 2.8 & 3.3 & 3.8 & 4.5 & 2.2 & 2.4 & 3.5 & 3.4\\
 & -0.3 & 4.5 & 4.9 & 3.3 & 6.8 & 2.3 & 2.5 & 3.7 & 4.6 & 2.6 & 3.5 & 2.0 & 3.2\\
C & 0.0 & 6.0 & 5.8 & 7.5 & 8.5 & 3.9 & 5.2 & 6.0 & 7.1 & 4.0 & 4.6 & 4.5 & 6.1\\
 & 0.3 & 3.4 & 3.9 & 2.7 & 4.8 & 1.9 & 3.1 & 3.7 & 4.0 & 2.4 & 1.8 & 3.1 & 3.8\\
 & 0.6 & 3.7 & 3.4 & 3.1 & 4.3 & 2.5 & 3.2 & 3.9 & 5.6 & 2.8 & 2.8 & 4.9 & 6.0\\
 & 0.9 & 4.5 & 4.8 & 5.4 & 4.8 & 3.8 & 4.7 & 7.1 & 5.0 & 3.0 & 3.3 & 5.6 & 10.4\\
\cmidrule(l{3pt}r{3pt}){3-6} \cmidrule(l{3pt}r{3pt}){7-10} \cmidrule(l{3pt}r{3pt}){11-14}
\end{tabular}
\end{table}

As for the power of the test, the presence of high levels of measurement noise impinge negatively upon it so that larger sample sizes are required to compensate for this. The study is reported in Section~\ref{Supp:merr} of the Supplementary Material.
%
\section{Testing and Modelling the Italian Strikes Time Series}\label{sec:real}
%
\begin{figure}
	\centering
    \includegraphics[width=0.45\linewidth,keepaspectratio]{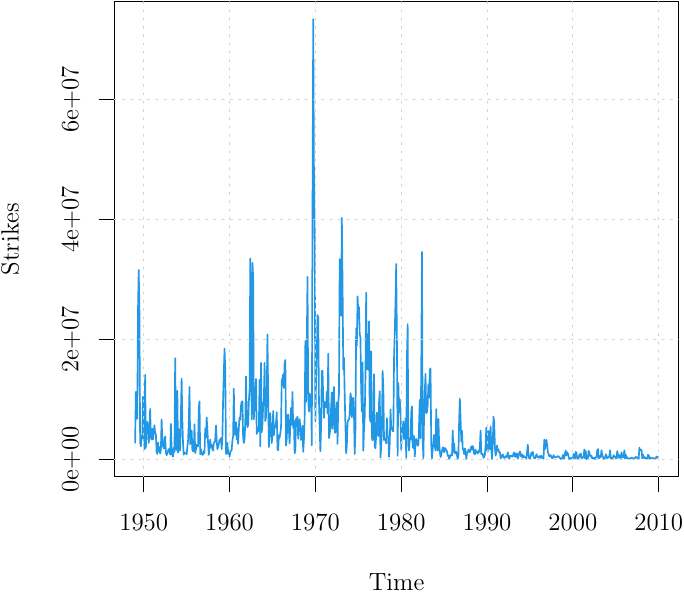}
	\includegraphics[width=0.45\linewidth,keepaspectratio]{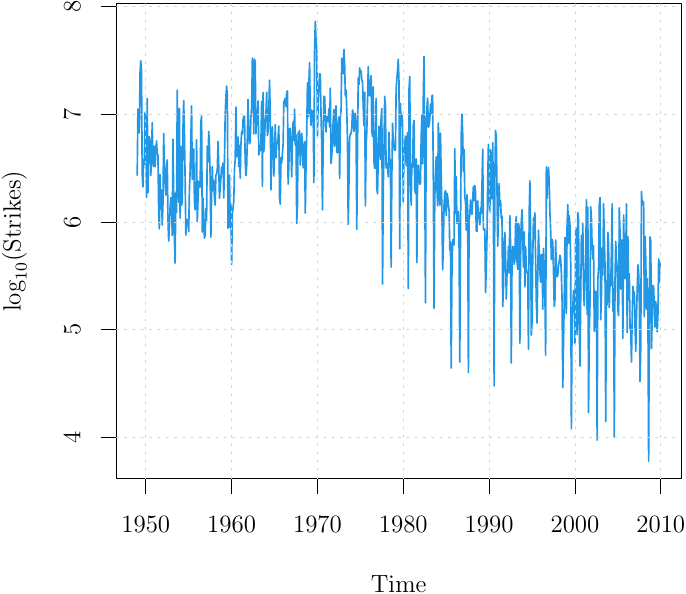}
	\caption{Time series of the Italian Strikes from January 1949 to December 2009. (Left) raw time series. (Right) $\log_{10}$-transformed series.}\label{fig:1}
\end{figure}
In this section, we study the dynamics of the monthly series of hours not worked due to strikes in Italy between January 1949 and December 2009 ($n=732$). The data were obtained from the Italian Statistical Institute (ISTAT). Despite their importance, to the best of our knowledge, this is the first time that these data are used within a macroeconometric approach. Note that, also for Italy, ISTAT has suspended the collection of monthly data from 2009 onwards.
The time plot is characterized by cyclical oscillations and a structural change in variance starting from the mid-1980s, see Figure~\ref{fig:1} (left). In the Italian history, the amplitude of such oscillation seems to reduce starting from the 1980s, without getting back to the previous variability in successive years. This can be explained by a series of historical occurrences such as the increase in the workers' opportunity cost associated to the decision to strike, which reduced the incentive to strike for economic reasons, e.g. wage and/or contractual claims. 
\par
We consider the logarithm (in base 10) as a variance-stabilising transformation and show the result in Figure~\ref{fig:1} (right). The month plot is shown in Figure~\ref{fig:2}  of the Supplementary Material. The series has seasonal oscillations and August is the month where the strikes undergo a consistent drop, followed by September, January, and December, where the phenomenon is less pronounced. August is a vacation month in Italy, and September is a post-vacation month, while January and December are characterized by less working days due to Christmas holidays. The series also presents some seasonality in the minima and the maxima, due to the strikes being more likely to be observed in those months when they are more effective. As strikes are costly for both firms and workers, the seasonality of economic activity, as well as the economic cycle, make the strike strategy more effective during the high seasons or expansionary phases while discouraging workers during low seasons or recessionary phases.
\par
\begin{figure}
\centering \includegraphics[width=0.8\linewidth,keepaspectratio]{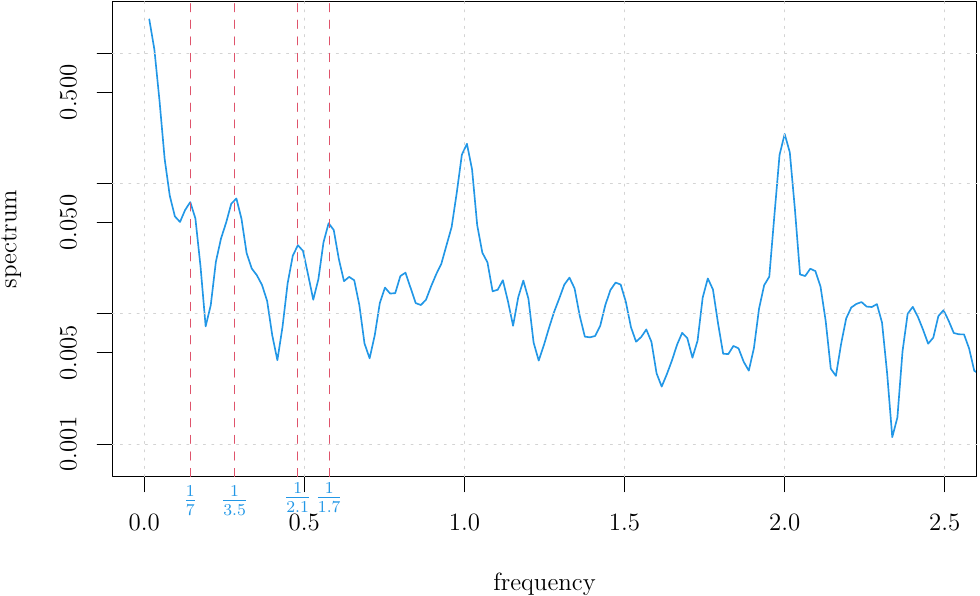}
\caption{Spectral density function, up to a frequency of 2.5 (cycles per year) of the Italian Strikes (in $log_{10}$) from January 1949 to December 2009. The periods corresponding to the dominant peaks are added in blue.}\label{fig:5}
\end{figure}

The autocorrelation of the series decays slowly hinting at the presence of a trend and a strong and seemingly nearly integrated seasonal component.\footnote{The correlograms are reported in Supplementary Material, Section~\ref{Supp:Graph}.} 
The spectral density function (smoothed periodogram with a Daniell window) of Figure~
\ref{fig:5} shows the main periodicities of the series. Besides the yearly periodicity, the 7-year peak together with its harmonics (3.5, 1.7, and 0.85 years) are related to the business cycle. The 2.1-year periodicity could be a first indication of a non-linear dynamics emerging as a resonance (non-trivial combination of the main frequencies). This periodicity is also in line with the existence of a cycle in Italian strikes recalled by \citet{franzosi1980} based on the spectral analysis of this same series, using data predating the 1980s. Indeed, strike activity concentrates at the expiration of the contracts, whose average duration is, 2-3 years \citep{myers2016}.
Labour strikes are a tool for bargaining better working conditions, both for those who strike and for those who do not, and this poses a threat to firms \citep{barrett1989}. Strikes have an upper bound in the number of hours they can be used, though, since above that level the tool can backfire and damage the workers themselves. This can be due, for example, to the firm being so damaged by the strikes that it has to reduce its production and hence lay off some workers. This implies that the series cannot present a unit root, be it regular or seasonal. Moreover, cointegrating relations are not expected. We assess this conjecture by applying a battery of unit root tests to the series of the Italian strikes and to the following covariates: \emph{salary}: Monthly index of salaries of industrial workers; \emph{price}:  Monthly index of consumer prices\footnote{All the series have been obtained from the Italian National Institute of Statistics (ISTAT).} Following the labor supply models, we have chosen these two economic variables because they represent a measure of the opportunity cost of strikes and one of the main economic motivations behind strikes for wage demands. The results are shown in Table~\ref{tab:ur1}. The first row presents the test $\bar MZ_{\alpha}^{\text{GLS}}$ as proposed by \cite{Per07}, which is essentially the same as the $MZ_{\alpha}^{\text{GLS}}$ test of \cite{Ng01} but the lag of the ADF regression is selected on OLS detrended data. The second row shows the results for the $\bar MP_{t}^{\text{GLS}}$, the modified feasible point optimal test (see also Eq.~(9) in \cite{Ng01}). The third row lists the GLS detrended version of the ADF test (denoted by ADF$^{\text{GLS}}$). Following \cite{Cha20b}, we have chosen them since they are the best performers among all those proposed in \cite{Ng01} and \citet{Per07}.\footnote{We have ported to R the original Gauss routines of \cite{Ng01}, which are available at \url{https://drive.google.com/open?id=0B-aG4lrQrBYsazN1RktHX2dfZkU}. The results for the remaining tests are similar and are available upon request.}  The last column contains the critical values of the null asymptotic distribution at the 5\% level. Clearly, none of the three tests manages to reject the null hypothesis of a unit-root in any of the series.
\begin{table}
\centering
\caption{Values of three unit root test statistics applied to the series of the Italian strikes plus the 3 covariates. The last column reports the 5\% critical values of the null distribution.}\label{tab:ur1}
\begin{tabular}{lrrrr}

  & \emph{strikes} &  \emph{salary} & \emph{price} & 5\% critical value\\
    \cmidrule(lr){2-5}
$\bar MZ_{\alpha}^{\text{GLS}}$    & -1.71 &  -1.23 & -6.14 & -21.30 \\
        $\bar MP_{t}^{\text{GLS}}$ & 52.92 &  70.04 & 14.83 &   5.48 \\
        ADF$^{\text{GLS}}$         & -1.08 &  -1.52 & -1.64 &  -2.91 \\
    \cmidrule(lr){2-5}
\end{tabular}
\end{table}
For this reason, we perform a pairwise cointegration analysis. We regress the series of strikes on the three covariates and apply the above tests to the residuals of the fitted models. The results are presented in Table~\ref{tab:ur2}. Again, none of the tests is able to reject the null hypothesis and this renders the whole analysis inconclusive in that all the series appear to be integrated but no cointegrating relationships can be found.

\begin{table}
\centering
\caption{Cointegration tests: application of the unit root tests to the residuals of the regression of the Italian strikes on the 3 covariates. The last column reports the 5\% critical values of the null distribution.}\label{tab:ur2}
 \begin{tabular}{lrrr}
 & \emph{salary} & \emph{price} & 5\% critical value\\
    \cmidrule(lr){2-4}
$\bar MZ_{\alpha}^{\text{GLS}}$    & -1.81 & -2.92 & -21.30\\
        $\bar MP_{t}^{\text{GLS}}$ & 50.21 & 31.19 &  5.48\\
        ADF$^{\text{GLS}}$         & -1.21 & -1.83 & -2.91\\
    \cmidrule(lr){2-4}
\end{tabular}
\end{table}


One possible reason for the above results is the lack of power of unit root tests against a non-linear alternative. This could be due to them not explicitly encompassing the nonlinearity within their specification. On the other hand, tests for a unit root against a threshold autoregressive alternative suffer from the presence of MA components and are severely biased, leading to over-rejecting. This is discussed in \cite{Cha20b,Cha24}, which solves the problem by proposing a unit root test where the null hypothesis entails an integrated MA against the alternative of a stationary threshold ARMA model, possessing a unit root regime. The application of such a test to the time series of Italian strikes produces a test statistic equal to 13.95, with an associated (heteroskedastic robust) wild bootstrap $p$-value of 0.012, and this points to an alternative explanation of the strikes dynamics based upon a regime-switching mechanism.
\par

\begin{figure}
\centering
\includegraphics[width=0.8\linewidth,keepaspectratio]{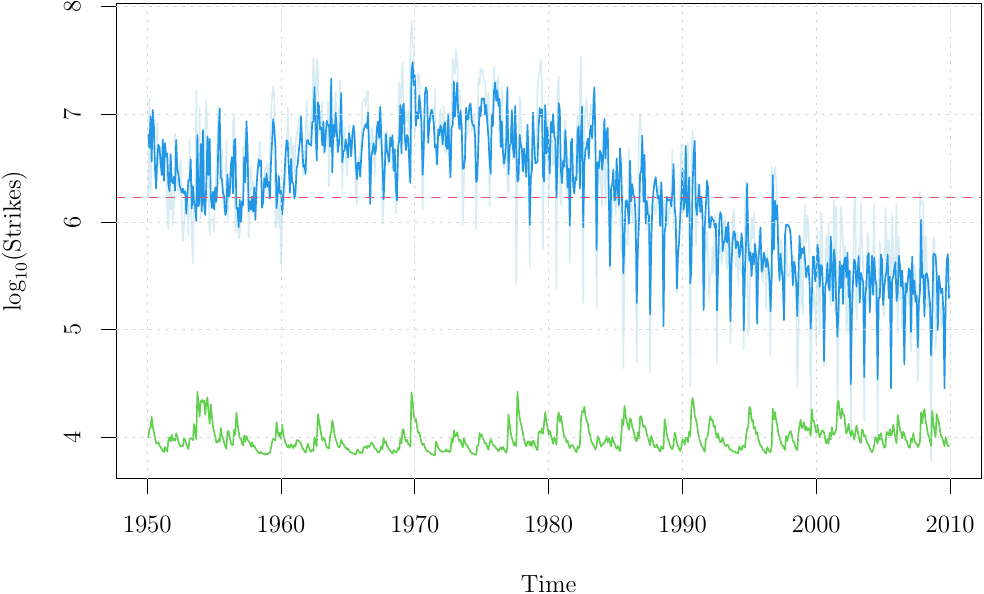}
\caption{Time series of the Italian Strikes (light blue) with the fitted series from the TARMA-GARCH model (blue) and the estimated conditional standard deviations $\hat h^{1/2}_t$ (green). The estimated threshold is the dashed red line.}\label{fig:16}
\end{figure}

In Section~\ref{Supp:GARCH} of the Supplementary Material, we adopt a linear modelling approach based on a two-step ARIMA-GARCH fit. The results show that both regular and seasonal differences are needed and this is not consistent with stylized economic facts and difficult to interpret. Moreover, the residual analysis hints at the presence of unaccounted non-linear dependence (see Figure~\ref{fig:9} of the Supplementary Material). For these reasons, we specify a TARMA-GARCH model and test it against the ARMA-GARCH specification by using our novel sup-Lagrange Multiplier test to take into account the conditional heteroskedasticity. We use the consistent Hannan-Rissanen criterion to select the order of the ARMA model to be tested \citep{Han82}. The maximum order tested is the ARMA(12,12) and the procedure used selects the ARMA(1,1) model. 
As for the order of the GARCH model specification we select the GARCH(1,2) on the basis of previous investigations. Hence we test the ARMA(1,1)-GARCH(1,2) against the TARMA(1,1)-GARCH(1,2) specification with $d=1$. We obtained a value of the supLM statistic equal to 25.668 (threshold = 6.273). Given that the critical value at 1\% level results 17.65 (from Table~I of \cite{And03} with $\pi_0=.20$), our asymptotic sLMg test rejects and points to a significant threshold effect either in the intercept and/or at lag 1, so the results of the test corroborate the hypothesis that the dynamic of strikes is governed by a regime-switching dynamics with conditional heteroskedasticity. We propose the following two-stage TARMA-GARCH model:

\begin{align}\label{eq:TG}
	X_t &=
	\begin{cases}
		\phi_{1,0}+\phi_{1,1}X_{t-1}+\phi_{1,12}X_{t-12}+\theta_{1,3}\eps_{t-3}+ \theta_{1,12}\eps_{t-12}+ \theta_{1,13}\eps_{t-13} + \eps_t & \text{if } X_{t-1}\leq r \\
		\phi_{2,0}+\phi_{2,1}X_{t-1}+\phi_{2,12}X_{t-12}+\theta_{2,1}\eps_{t-1}+ \theta_{2,3}\eps_{t-3} + \eps_t, & \text{if } X_{t-1}> r.
	\end{cases}\nonumber \\
	\eps_t&=\sqrt{h_t}z_t; \quad h_t=a_0 + a_1\eps^2_{t-1} + b_1 h_{t-1}.
\end{align}
\noindent
We adopt a two-stage estimation approach since the existing results for TARMA models include least squares estimators for which consistency and asymptotic normality have been established \citep{Gia21, Li11b}. In this way, we can adopt a maximum likelihood approach for estimating the GARCH part and exploit mature and reliable existing implementations such as those of the R package \texttt{rugarch} \citep{Gha20}. Equation~\eqref{tarma.fit} reports the estimated coefficients of the model, with standard errors in parentheses below each coefficient. The threshold is estimated to be 6.23, equivalent to roughly 1.7 million non-worked hours due to strikes in a month.
\begin{align}
X_t &=
\begin{cases}
\underset{(0.15)}{0.42} +
\underset{(0.03)}{0.07} X_{t-1} +
\underset{(0.03)}{0.85} X_{t-12} +
\underset{(0.04)}{0.09} \eps_{t-3} -
\underset{(0.05)}{0.51} \eps_{t-12} -
\underset{(0.05)}{0.19} \eps_{t-13} +
\eps_t, & \text{if } X_{t-1} \leq 6.23\\
\underset{(0.48)}{0.77} +
\underset{(0.08)}{0.41} X_{t-1} +
\underset{(0.05)}{0.47} X_{t-12} +
\underset{(0.07)}{0.25} \eps_{t-1} +
\underset{(0.05)}{0.05} \eps_{t-3} +
\eps_t, & \text{if } X_{t-1} > 6.23
\end{cases} \label{tarma.fit}\\
	\eps_t &=\sqrt{h_t}z_t; \quad
	 h_t   =\underset{(0.09)}{0.20} + \underset{(0.04)}{0.10}\eps^2_{t-1} + \underset{(0.10)}{0.70} h_{t-1}. \label{garch.fit}
\end{align}
\noindent
The estimated TARMA model fulfils the sufficient conditions of ergodicity and invertibility as discussed in \cite{Cha19}. Figure~\ref{fig:16} shows the series (light blue line) together with the fitted values from the TARMA model (blue line), the estimated conditional standard deviation $\hat h^{1/2}_t$ from the GARCH(1,1) fit (green line), and the estimated threshold ($\hat r = 6.23$, dashed red line). The TARMA specification for the conditional mean has a lower regime which is characterized by a persistent seasonality with a clear threshold effect at lag 12. The MA(3) parameters are not significant but play a role in providing a solution that mimics the observed dynamics. Note that the upper regime contains no seasonal MA lags and this is a strong indication of the presence of a threshold effect in the moving average model, something that conventional TAR models are not able to reproduce or, at best, can only reproduce partially by incorporating higher order lags. Clearly, the dynamics reacts differently to shocks depending upon the regime. The threshold identifies the time corresponding to the structural break in the variance of the series and splits it into two periods. Indeed, from the mid-eighties onwards the series belongs to the lower regime where the dynamics is much more characterized by seasonality. The series of the conditional standard deviation shows both clusters of volatility and asymmetric peaks, and appears to change starting from late 1990s.
\par
\begin{figure}
	\centering
\includegraphics[width=0.4\linewidth,keepaspectratio]{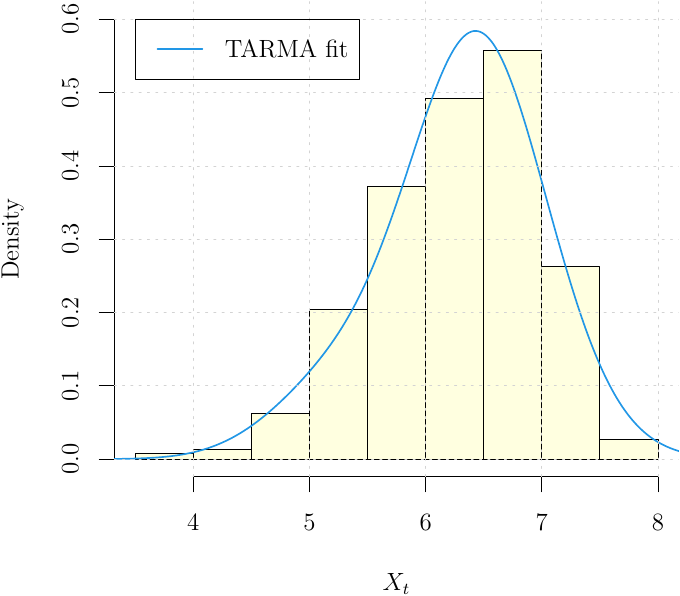}
\includegraphics[width=0.4\linewidth,keepaspectratio]{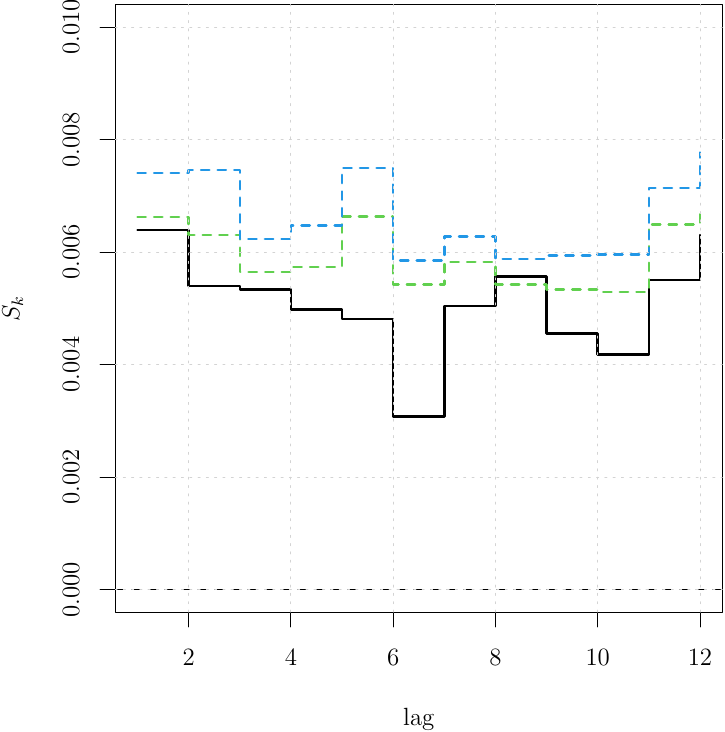}
\caption{(left) Histogram of the time series of the Italian strikes. The smooth blue curve is the density estimate based on 100k data simulated from the fitted TARMA model. (right) Entropy-based metric up to lag 12, computed on the residuals of the TARMA-GARCH model of Eq.~(\ref{eq:TG}). The dashed lines correspond to bootstrap rejection bands at levels 95\% (green) and 99\% (blue) under the null hypothesis of linear serial dependence.}\label{fig:10}
\end{figure}

The parameter estimates for the GARCH(1,1) part of the model are shown in Equation~\eqref{garch.fit}. The estimates indicate a non-negligible heteroskedastic structure in the innovations. Also in this case, the estimated model results stationary and ergodic and its adequacy is assessed in the Supplementary materials \ref{Supp:Diag}, which also reports model diagnostics. 
The fit is able to reproduce the strong asymmetric behaviour of the series: the histogram of the data in reported in Figure~\ref{fig:10} (left), where we superimposed a kernel density estimate over a simulated trajectory of 100k observations from the fitted model (blue line). This is also witnessed by the Kolmogorov-Smirnov two-sample test, computed between the distribution of the observed data and of that of the simulated series ($p$-value = 0.08). Moreover, we test the residuals of the model for the presence of non-linear serial dependence, up to lag 12, with the entropy metric $S_{\rho}$, see Figure~\ref{fig:10} (right). Since no lags exceed the bootstrap rejection band we can be confident that the model managed to capture the underlying non-linear features of the series. Finally, in Figure~\ref{fig:15} of the Supplementary Material we show that there is no unaccounted cross-dependence between the residuals of the fitted model and the covariates \emph{salary} and \emph{price}, introduced at the beginning of the section.

\par
The existence of a threshold-type dynamics is consistent with the underlying theory, see e.g. \citet{granovetter1978}, where individuals make choices that are influenced by the behaviour of other individuals. Also, \citet{bursztyn2021} stress the importance of the activity of people within their social network when choosing to take part in a protest. The existence of strong ties between individuals is important for the persistence of the protests and the clustering effect. 
As for the structural change in the variance, we note that the strengthening of the government and the exclusion of unions from the process of reforming policies has downplayed collective strikes as a political and social bargaining tool \citep{hamann2013}. \citet{godard2011} argues that the change in the dynamics of strikes observed since the 1980s can be due to four different reasons: $i)$ they have been diverted into alternative forms of conflict; $ii)$ the capitalist system managed to reduce the number of citizens who disapprove of the economic and political system itself, or at least it managed to reduce their will to act upon it;  $iii)$  the conflict underwent a transformation and has become more deeply embedded and linked to general behaviour within and outside the workplace, such as cynicism and escapism, but also depression; $iv)$ conflict has become dormant. While some of these reasons appear more convincing than others, their multifactorial interaction could have played a role in explaining the observable shift. In addition, other possible reasons, partially overlapping the aforementioned, are the reduction of the union density, the birth of autonomous and non-political trade unions, and the creation of local agreements, three events that characterized Italy after the 1970s and that are particularly relevant in the 1990s and 2000s, see also \citet{giangrande2021}.
\par

\section{Conclusions}\label{sec:conc}

Our proposal, that compares the ARMA-GARCH versus the TARMA-GARCH specifications, allows us to test for nonlinearity in the conditional mean, without being affected by heteroskedasticity, making it possible to assess the two types of nonlinearity separately. Since threshold ARMA models can accommodate a wide range of complex dynamics, we expect the test to have power against most types of nonlinearity in the conditional mean, besides threshold models. For this reason, it can be used as an omnibus explorative nonlinearity test with good properties of robustness against heteroskedasticity. This aspect will be further explored in future investigations.
\par
The results point to a significant threshold effect in the series of Italian strikes, which appears to be governed by a regime-switching dynamics with conditional heteroskedasticity. The analysis of the strikes offers some evidence that TARMA models provide a flexible and parsimonious specification that describes some of its empirical non-linear stylized facts. Our findings are coherent with both features of the history of the Italian labour market, such as the decrease in the social turmoil after the first half of the 1980s \citep{godard2011}, and the social aspects of unrest and labour markets in general, based on the threshold model by \citet{granovetter1978} and also in line with more recent studies of the impact of social ties and the existence of social networks able to trigger political engagement and participation to protest \citep{bursztyn2021}.


\bibliographystyle{plainnat}
\bibliography{Strikes_V12}

\section*{Supplementary Material}

The Supplementary Material contains all the proofs, additional Monte Carlo results and further results on the analysis of the time series of Italian strikes. The routines for testing and modelling are included in the \texttt{R} package \href{https://cran.mirror.garr.it/CRAN/web/packages/tseriesTARMA/readme/README.html}{\texttt{tseriesTARMA}}.

\setcounter{section}{0}
\setcounter{table}{0}
\setcounter{figure}{0}
\renewcommand\thesection{S.\arabic{section}}
\renewcommand\thetable{S.\arabic{table}}
\renewcommand\thefigure{S.\arabic{figure}}

\section{Proofs}\label{SMsec:proofs}

\subsubsection*{Proof of Proposition~\ref{prop:asymptotics}}
\textbf{Part (i)} By deploying the same argument in \citet{Gor23}, it is possible to prove that $n^{-1}\hat{\mathcal{I}}_n(r)=n^{-1}\mathcal{I}_n(\bEta_0,r)+o_p(1)$. The ergodicity of $\{X_t\}$ implies that, for each $r$, $n^{-1}\mathcal{I}_n(\bEta_0,r)$ converges in probability to $E[\partial^2\ell_n(\bEta_0,r)/\partial\bPsi\partial\bPsi^\intercal]$. On the other hand, it holds that:
\begin{align*}
&\frac{\partial^2\ell_n(\bEta_0,r)}{\partial\bPsi\partial\bPsi^\intercal}\\
&= \frac{1}{h_t}\left(\frac{\partial\eps_t(\bEta_0,r)}{\partial\bPsi}\right)\left(\frac{\partial\eps_t(\bEta_0,r)}{\partial\bPsi}\right)^\intercal - \frac{\eps_t}{h_t^2} \left(\frac{\partial h_t(\bEta_0,r)}{\partial\bPsi}\right)\left(\frac{\partial\eps_t(\bEta_0,r)}{\partial\bPsi}\right)^\intercal\\
& + \frac{\eps_t}{h_t}\frac{\partial^2\eps_t(\bEta_0,r)}{\partial\bPsi\partial\bPsi^\intercal} - \frac{\eps_t}{h_t^2} \left(\frac{\partial \eps_t(\bEta_0,r)}{\partial\bPsi}\right)\left(\frac{\partial h_t(\bEta_0,r)}{\partial\bPsi}\right)^\intercal\\
& + \frac{\eps_t^2}{h_t^3}\left(\frac{\partial h_t(\bEta_0,r)}{\partial\bPsi}\right)\left(\frac{\partial h_t(\bEta_0,r)}{\partial\bPsi}\right)^\intercal -  \frac{1}{2}\frac{1}{h_t^2}\left(\frac{\partial h_t(\bEta_0,r)}{\partial\bPsi}\right)\left(\frac{\partial h_t(\bEta_0,r)}{\partial\bPsi}\right)^\intercal\\
& + \frac{1}{2} \left(\frac{\eps_t^2}{h_t^2}-\frac{1}{h_t}\right)\frac{\partial^2 h_t(\bEta_0,r)}{\partial\bPsi\partial\bPsi^\intercal}.
\end{align*}
By combining the law of iterated expectations with $E[\eps_t|\mathcal{F}_{t-1}]=0$ and $E[\eps^2_t|\mathcal{F}_{t-1}]=h_t$ we have that, for each $r$,
$$\left\|M_n(r)\right\|=o_p(1),\qquad\text{where } M_n=\frac{\mathcal{I}_n(\bEta_0,r)}{n}-\Lambda(r).$$
In the following, we prove that the result holds uniformly, i.e.:
$$\sup_{r\in[r_l,r_u]}\left\|M_n(r)\right\|=o_p(1).$$
To this aim fix $\kappa_1<\kappa_2$ and consider a grid $\kappa_1=r_0<r_1<\ldots< r_m=\kappa_2$ with equal mesh size, i.e. $r_i-r_{i-1}\equiv c$, for some $c>0$.
It holds that
$\sup_{r \in [r_{i-1}, r_i]}\|M_n(r)-M_n(r_{i-1})\|\le C_n$,
for all $i$. Moreover, $E(C_n)\to 0 $ as  $c\to 0$. Because for any $r \in [\kappa_1,\kappa_2]$, there exists an $i$ such that $r_{i-1} \le r \le r_{i}$ and hence
$M_n(r)=M_n(r)- M_n(r_{i-1})+M_n(r_{i-1})$  and
$\sup_{r\in [\kappa_1,\kappa_2]}\|M_n(r)\|\le \max_{i=0,\ldots,m} M_n(r_i)+C_n.$
The proof is complete since for fixed $m$,
$$\max_{i=0,\ldots,m} M_n(r_i) \to 0 \text{ in probability}$$
  and $E(C_n)\to 0$ as $c\to 0$ in probability.
  \par\noindent
\textbf{Part (ii)} Within this proof, all the $o_p(1)$ terms hold uniformly on $r\in[r_L,r_U]$. We need to prove that:
\begin{align*}
\sup_{r\in[r_L,r_U]}&\left\|\frac{1}{\sqrt{n}} \frac{\partial\hat\ell_n(r)}{\partial\boldsymbol\Psi_2}-\left(\nabla_{n,2}(r)- \Lambda_{21}(r)\Lambda_{11}^{-1} \nabla_{n,1}\right)\right\| = o_p(1).
\end{align*}
Since $\sqrt{n}(\hat{\bPsi}_1-\bPsi_{0,1})=\Lambda_{11}^{-1}n^{-1/2}\partial\ell_n/ \partial\bPsi_1+o_p(1)$ and $\Lambda_{21}(r)=O_p(1)$ uniformly in $r\in[r_L,r_U]$, then it is sufficient to prove that
\begin{align*}
\sup_{r\in[r_L,b_U]} \left\|\frac{1}{\sqrt{n}}\frac{\partial\hat\ell_n(r)}{\partial\bPsi_2} - \frac{1}{\sqrt{n}}\frac{\partial\ell_n(r)}{\partial\bPsi_2}+\Lambda_{21}(r)\sqrt{n}(\hat{\bPsi}_1 -\bPsi_{0,1})\right\|
=o_p(1).
\end{align*}
which holds by using the same arguments developed in Theorem 2.2 in \citet{Li11}.
\subsubsection*{Proof of Theorem~\ref{th:FCLT}}
To prove the FCLT we follow two steps: $(i)$ we verify that $Q_n(r)$ converges to $\xi_n(r)$ in terms of finite distributions and $(ii)$ we show the asymptotic equicontinuity  of $Q_n(r)$. The first step is readily implied by combining the central limit theorem and the Cramer-Wold device ergo the theorem will be proved upon showing the tightness of the score vector $\nabla_n(r)$ componentwise. For the sake of presentation and without loss of generality, we detail the case $p=q=u=v=1$ and prove the tightness of the following process:
\begin{align*}
G_n(r)&=\frac{1}{\sqrt{n}}\frac{\partial\ell_n(r)}{\partial\varphi_1}\\
&=-\frac{1}{\sqrt{n}}\sum_{t=1}^{n}\frac{\eps_t}{h_t}\frac{\partial\eps_t(\bEta_0,r)}{\partial\varphi_1} +\frac{1}{2}\frac{1}{\sqrt{n}}\sum_{t=1}^{n}\frac{1}{h_t}\left(\frac{\eps_t^2}{h_t}-1\right)\frac{\partial h_t(\bEta_0,r)}{\partial\varphi_1}\\
&=\frac{1}{\sqrt{n}}\sum_{t=1}^{n}\frac{\eps_t}{h_t} \sum_{\iota=0}^{t-1}\theta_1^\iota X_{t-1-\iota} I(X_{t-1-\iota}\leq r)\\
&+\frac{a_1}{\sqrt{n}}\sum_{t=1}^{n}\frac{1}{h_t}\left(1-\frac{\eps_t^2}{h_t}\right)\sum_{\tau=0}^{t-1}b_1^\tau \eps_{t-1-\tau} \sum_{\iota=0}^{t-2-\tau}\theta_1^\iota X_{t-2-\tau-\iota} I(X_{t-2-\tau-\iota}\leq r).
\end{align*}

For each $r,s\in\mathds{R}$, define
\begin{align*}
g_t^{(1)}(r,s)&= \frac{\eps_t}{h_t} \sum_{\iota=0}^{t-1}\theta_1^\iota X_{t-1-\iota} I(s< X_{t-1-\iota}\leq r),\\
g_t^{(2)}(r,s)&= \frac{a_1}{h_t}\left(1-\frac{\eps_t^2}{h_t}\right)\sum_{\tau=0}^{t-1}b_1^\tau \eps_{t-1-\tau} \sum_{\iota=0}^{t-2-\tau}\theta_1^\iota X_{t-2-\tau-\iota} I(s< X_{t-2-\tau-\iota}\leq r).
\end{align*}
Clearly,
$$G_n(r) - G_n(s)= \frac{1}{\sqrt{n}}\sum_{t=1}^{n}\left\{g_t^{(1)}(r,s)+g_t^{(2)}(r,s)\right\}.$$
By using the same approach of \citet{Won97} it suffices to prove that there exists a constant $\mathrm{C}$ such that
\begin{equation}\label{eq:main_tight}
  E\left[\sup_{|r-s|<\delta}\left|g_t^{(1)}(r,s)+g_t^{(2)}(r,s)\right|^2\right]\leq \mathrm{C}\delta,
\end{equation}
which follows upon proving that there exist two constants, say $\mathrm{C}_1$ and $\mathrm{C}_2$, such that
\begin{align}
&E\left[\left|g_t^{(1)}(r,s)\right|^2\right] \leq \mathrm{C}_1(r-s),\label{eq:main_tight1}\\
&E\left[\left|g_t^{(2)}(r,s)\right|^2\right] \leq \mathrm{C}_2(r-s).\label{eq:main_tight2}
\end{align}
In the following we use $\mathcal{C}$ to refer to a generic constant that can change across lines. In order to prove (\ref{eq:main_tight1}) note that:
$$E\left[\left.\frac{\eps_t^2}{h_t^2}\right|\mathcal{F}_{t-1}\right]=\frac{1}{h_t}\leq \frac{1}{a_0}.$$
Hence (\ref{eq:main_tight1}) is verified by using the law of iterated expectations, Jensen's inequality and the fact that $|\theta_1|<1$:
\begin{align*}
E\left[\left|g_t^{(1)}(r,s)\right|^2\right]&= E\left[\left\{\sum_{\iota=0}^{t-1}\theta_1^\iota X_{t-1-\iota} I(s< X_{t-1-\iota}\leq r)\right\}^2 E\left[\left.\frac{\eps_t^2}{h_t^2}\right|\mathcal{F}_{t-1}\right] \right]\\
&\leq\frac{1}{a_0(1-|\theta_1|)}  \sum_{\iota=0}^{t-1}|\theta_1|^\iota E\left[ X^2_{t-1-\iota} I(s< X_{t-1-\iota}\leq r)\right]\\
&\leq \mathrm{C}_1 (r-s).
\end{align*}
As concerns (\ref{eq:main_tight2}) note that:
\begin{align*}
&E\left[ \left.\left|1-\frac{\eps_t^2}{h_t}\right|^2\right|\mathcal{F}_{t-1}\right]\leq E\left[ \left.4\left(1+\frac{\eps_t^4}{h_t^2}\right)\right|\mathcal{F}_{t-1}\right]\\
&=4E\left[ \left.1+\frac{z_t^4h_t^2}{h_t^2}\right|\mathcal{F}_{t-1}\right] = 4(1+E[z_t^4])<\infty.
\end{align*}
The law of iterated expectations and Jensen's inequality imply that
\begin{align}
&E\left[\left|g_t^{(2)}(r,s)\right|^2\right]= E\left[\left|\frac{a_1}{h_t}\left(1-\frac{\eps_t^2}{h_t}\right)\sum_{\tau=0}^{t-1}b_1^\tau \eps_{t-1-\tau} \sum_{\iota=0}^{t-2-\tau}\theta_1^\iota X_{t-2-\tau-\iota} I(s< X_{t-2-\tau-\iota}\leq r)\right|^2\right] \nonumber\\
&=a_1^2 E\left[\frac{1}{h_t^2}\left|\sum_{\tau=0}^{t-1}b_1^\tau \eps_{t-1-\tau} \sum_{\iota=0}^{t-2-\tau}\theta_1^\iota X_{t-2-\tau-\iota} I(s< X_{t-2-\tau-\iota}\leq r) \right|^2E\left[ \left.\left|1-\frac{\eps_t^2}{h_t}\right|^2\right|\mathcal{F}_{t-1}\right]\right] \nonumber\\
&\leq \mathcal{C} E\left[\frac{1}{h_t^2}\left|\sum_{\tau=0}^{t-1}b_1^\tau \eps_{t-1-\tau} \sum_{\iota=0}^{t-2-\tau}\theta_1^\iota X_{t-2-\tau-\iota} I(s< X_{t-2-\tau-\iota}\leq r) \right|^2\right] \nonumber\\
&\leq \mathcal{C} \sum_{\tau=0}^{t-1}|b_1|^\tau \sum_{\iota=0}^{t-2-\tau}|\theta_1|^\iota E\left[ \eps^2_{t-1-\tau}X^2_{t-2-\tau-\iota}I(s< X_{t-2-\tau-\iota}\leq r)\right]\nonumber\\
&= \mathcal{C} \sum_{\tau=0}^{t-1}|b_1|^\tau \sum_{\iota=0}^{t-2-\tau}|\theta_1|^\iota E\left[ X^2_{t-2-\tau-\iota}I(s< X_{t-2-\tau-\iota}\leq r) E\left[\left.\eps^2_{t-1-\tau}\right|\mathcal{F}_{t-2-\tau}\right]\right]\nonumber\\
&= \mathcal{C} \sum_{\tau=0}^{t-1}|b_1|^\tau \sum_{\iota=0}^{t-2-\tau}|\theta_1|^\iota E\left[h_{t-1-\tau} X^2_{t-2-\tau-\iota}I(s< X_{t-2-\tau-\iota}\leq r)\right]\label{eq:main21}.
\end{align}
For each $k\in\mathds{N}$, $k\neq 0$, consider $E[h_{t-1-\tau} X^2_{t-k}I(s< X_{t-k}\leq r)]$. Since, with the convention that $\prod_{i=1}^{0}\cdot=1$,
$$h_t=a_0\sum_{j=0}^{k-1}\prod_{i=1}^j\left(a_1z^2_{t-i}+ b_1\right) + \prod_{i=1}^{k}\left(a_1z^2_{t-i}+ b_1\right)h_{t-k}$$
it holds that
\begin{align}
&E[h_{t-1-\tau} X^2_{t-k}I(s< X_{t-k}\leq r)]\nonumber\\
&=E\left[\left\{a_0\sum_{j=0}^{k-1}\prod_{i=1}^j\left(a_1z^2_{t-i}+ b_1\right) + \prod_{i=1}^{k}\left(a_1z^2_{t-i}+ b_1\right)h_{t-k}\right\} X^2_{t-k}I(s< X_{t-k}\leq r)\right]\nonumber\\
&= E\left[a_0\sum_{j=0}^{k-1}\prod_{i=1}^j\left(a_1z^2_{t-i}+ b_1\right) X^2_{t-k}I(s< X_{t-k}\leq r)\right]\label{eq:main21a}\\
&+E\left[ \prod_{i=1}^{k}\left(a_1z^2_{t-i}+ b_1\right)h_{t-k} X^2_{t-k}I(s< X_{t-k}\leq r)\right]\label{eq:main21b}.
\end{align}
We prove separately that (\ref{eq:main21a}) and (\ref{eq:main21b}) are bounded by $\mathcal{C}(r-s)$. Since $E[z^2_{t_1}X^2_{t_2}]=E[z^2_{t_1}]E[X^2_{t_2}]$ for each $t_1>t_2$, and $E[z^2_{t}]=1$ for each $t$ (\ref{eq:main21a}) equals
\begin{align*}
&E\left[a_0\sum_{j=0}^{k-1}\prod_{i=1}^j\left(a_1z^2_{t-i}+ b_1\right) X^2_{t-k}I(s< X_{t-k}\leq r)\right]\\
&=a_0\sum_{j=0}^{k-1}\prod_{i=1}^j\left(a_1+ b_1\right)E\left[X^2_{t-k}I(s< X_{t-k}\leq r)\right]\\
&\leq \mathcal{C}(r-s).
\end{align*}
Analogously, (\ref{eq:main21b}) results
\begin{align*}
&E\left[ \prod_{i=1}^{k}\left(a_1z^2_{t-i}+ b_1\right)h_{t-k} X^2_{t-k}I(s< X_{t-k}\leq r)\right]\\
&=(a_1+b_1)^k E\left[h_{t-k}X^2_{t-k}I(s< X_{t-k}\leq r)\right]
\end{align*}
We claim that
\begin{equation}\label{eq:claim}
  E\left[h_{t-k}X^2_{t-k}I(s< X_{t-k}\leq r)\right]\leq \mathcal{C}(r-s).
\end{equation}
Therefore (\ref{eq:main21b}) is bounded by $(a_1+b_1)^k\mathcal{C}(r-s)$ whereas (\ref{eq:main21}) is bounded by
\begin{align*}
&\mathcal{C}\sum_{\tau=0}^{t-1}|b_1|^\tau \sum_{\iota=0}^{t-2-\tau}|\theta_1|^\iota\left\{(r-s)+(a_1+b_1)^{\iota+1}(r-s)\right\}\\
&\leq \mathrm{C}_2(r-s).
\end{align*}
Hence (\ref{eq:main_tight}) holds and the tightness of the process follows by using the same argument of \citet{Won97}. It remains to verify the Claim~\ref{eq:claim}. To this aim, define
\begin{align*}
\Upsilon_t^{(1)}=\phi_0+\phi_1X_t-\theta_1\eps_t\quad\text{and}\quad \Upsilon_t^{(2)}=\phi^2_0+\phi_1^2X^2_t+\theta^2_1\eps^2_t.
\end{align*}
Since $X_{t-k}=\Upsilon^{(1)}_{t-k-1}+\eps_{t-k}$ and $X^2_{t-k}\leq \mathcal{C}\left(\Upsilon^{(2)}_{t-k-1}+\eps^2_{t-k}\right)$, we have that
\begin{align}
&E\left[h_{t-k}X^2_{t-k}I(s< X_{t-k}\leq r)\right]\leq E\left[h_{t-k}\left(\Upsilon^{(2)}_{t-k-1}+\eps^2_{t-k}\right)I(s< \Upsilon^{(1)}_{t-k-1}+\eps_{t-k}\leq r)\right]\nonumber\\
&= \mathcal{C} E\left[h_{t-k}\Upsilon^{(2)}_{t-k-1}I(s< \Upsilon^{(1)}_{t-k-1}+\eps_{t-k}\leq r)\right]\nonumber\\
&+ \mathcal{C}E\left[h_{t-k}\eps^2_{t-k}I(s< \Upsilon^{(1)}_{t-k-1}+\eps_{t-k}\leq r)\right]\nonumber\\
&=\mathcal{C} E\left[h_{t-k}\phi_0^2I(s< \Upsilon^{(1)}_{t-k-1}+\eps_{t-k}\leq r)\right]\label{eq:claim1_a}\\
&+ \mathcal{C} E\left[h_{t-k}\phi_1^2X^2_{t-k-1}I(s< \Upsilon^{(1)}_{t-k-1}+\eps_{t-k}\leq r)\right]\label{eq:claim1_b}\\
&+ \mathcal{C} E\left[h_{t-k}\theta_1^2\eps^2_{t-k-1}I(s< \Upsilon^{(1)}_{t-k-1}+\eps_{t-k}\leq r)\right]\label{eq:claim1_c}\\
&+ \mathcal{C}E\left[h_{t-k}\eps^2_{t-k}I(s< \Upsilon^{(1)}_{t-k-1}+\eps_{t-k}\leq r)\right]\label{eq:claim2}
\end{align}
We prove that (\ref{eq:claim1_a}) -- (\ref{eq:claim2}) are bounded by $\mathcal{C}(r-s)$. In this respect, we show that
\begin{equation}\label{eq:main_claim}
  \Pr\left[\left.s< \Upsilon^{(1)}_{t-k-1}+\eps_{t-k}\leq r\right|\mathcal{F}_{t-k-1}\right]\leq \frac{\mathcal{C}}{\sqrt{a_0}}(r-s),
\end{equation}
 Indeed:
\begin{align*}
&\Pr\left[\left.s< \Upsilon^{(1)}_{t-k-1}+\eps_{t-k}\leq r\right|\mathcal{F}_{t-k-1}\right]\\
&= \Pr\left[\left.\frac{s- \Upsilon^{(1)}_{t-k-1}}{h_{t-k}^{1/2}}< \eps_{t-k}\leq \frac{r- \Upsilon^{(1)}_{t-k-1}}{h_{t-k}^{1/2}}\right|\mathcal{F}_{t-k-1}\right]\\
&=\int_{h_{t-k}^{-1/2}(s- \Upsilon^{(1)}_{t-k-1})}^{h_{t-k}^{-1/2}(r- \Upsilon^{(1)}_{t-k-1})}f_z(x)dx\\
&\leq \mathcal{C}\frac{1}{h_{t-k}^{1/2}}(r-s)\leq \frac{\mathcal{C}}{\sqrt{a_0}}(r-s).
\end{align*}
Routine algebra and (\ref{eq:main_claim})  imply that
\begin{align*}
&(\ref{eq:claim1_a}) \leq  \frac{\mathcal{C}}\sigma_\eps^2{\sqrt{a_0}}(r-s)\\
&(\ref{eq:claim1_b}) \leq  \frac{\mathcal{C}}\sigma_\eps^2{\sqrt{a_0}}(r-s)E[h_{t-k}X^2_{t-k-1}]\\
&(\ref{eq:claim1_c}) \leq  \frac{\mathcal{C}}\sigma_\eps^2{\sqrt{a_0}}(r-s)E[h_{t-k}\eps^2_{t-k-1}]\\
&(\ref{eq:claim2}) \leq  \frac{\mathcal{C}}\sigma_\eps^2{\sqrt{a_0}}(r-s)\\
\end{align*}
with $\sigma_\eps^2=E[h_t]$ being the (unconditional) variance of $\eps_t$ which is finite by Assumption A.3 Moreover it is not hard to prove that $E[h_{t-k}X^2_{t-k-1}]$ and $E[h_{t-k}\eps^2_{t-k-1}]$ are finite. For completeness we provide a sketch of th proof that $E[h_{t-k}X^2_{t-k-1}]<\infty$. Without loss of generality consider the case $k=1$. Combining the MA$(\infty)$-representation of the ARMA process, the ARCH$(\infty)$-representation of the GARCH process with Jensen's inequality, it holds that $E[h_tX^2_{t-1}]$ is bounded by
\begin{align*}
&\mathcal{C} E\left[\left\{\frac{a_0}{1+b_1}+a_1\sum_{j=0}^{t-1}b_1^j\eps^2_{t-1-j}\right\}\right.\\
&\left.\times\left\{\frac{\phi_0^2}{(1-|\theta_1|)^2}+\theta_1^2\sum_{i=0}^{t-2}|\phi_1|^i\eps^2_{t-2-i} + \sum_{i=0}^{t-2}|\phi_1|^i\eps^2_{t-1-i}\right\}\right]\\
&\leq \mathcal{C}\left\{\frac{a_0\phi_0^2}{(1-b_1)(1-|\phi_1|)^2} + \frac{a_0(1+\theta_1^2)}{(1-b_1)(1-|\phi_1|)}\right.\\
&\left. + \frac{a_0\phi_0^2\sigma_\eps^2}{(1-b_1)(1-|\phi_1|)^2} +  \frac{|\phi_1|}{(1-b_1|\phi_1|)(1-|\phi_1|(a_1+b_1))}\right\}
\end{align*}
which is finite and this complete the whole proof.


\renewcommand\thefigure{\thesection.\arabic{figure}}
\setcounter{figure}{0}
\renewcommand\thetable{\thesection.\arabic{table}}
\setcounter{table}{0}

\section{Supplementary Material: additional Monte Carlo results}\label{SMsec:MC}

\subsection{Size of the tests}

\begin{figure}[H]
\centering
\includegraphics[width=0.9\linewidth,keepaspectratio]{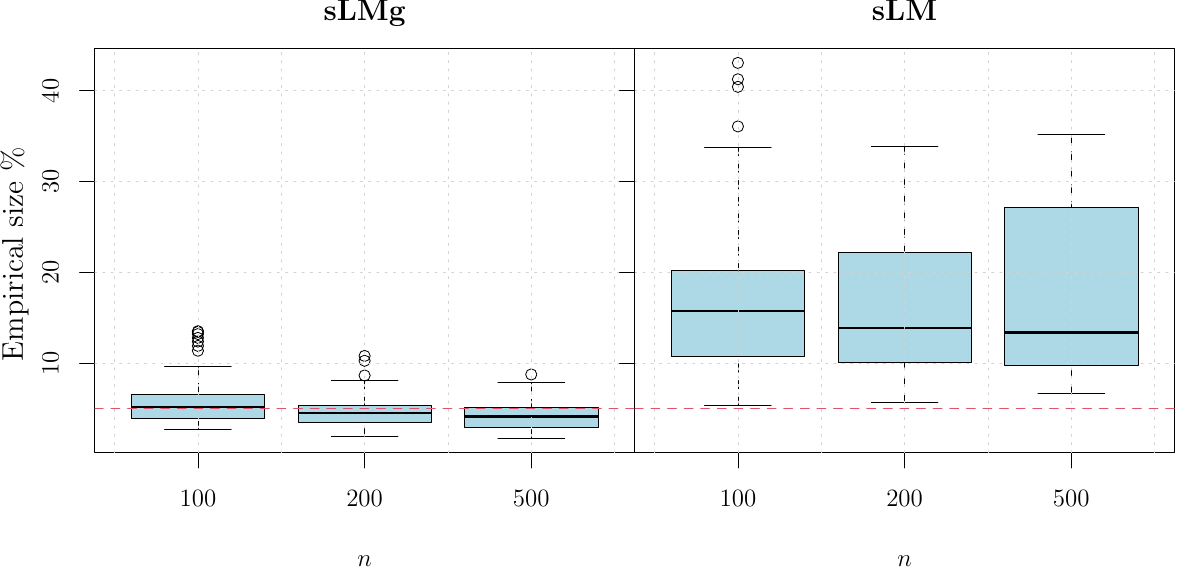}
 \caption{Empirical size (percent) at nominal level $\alpha=5\%$ for the ARMA(1,1)-GARCH(1,1) process of Eq.~(\ref{ARMAGARCHsim}). All the cases together.}\label{fig:size2}
\end{figure}

\begin{table}[H]
\centering
\small
\begin{tabular}{rrrrrrrr}
\multicolumn{1}{c}{$\phi_1$} & \multicolumn{1}{c}{$\theta_1$} & \multicolumn{2}{c}{$n=100$} & \multicolumn{2}{c}{$n=200$} & \multicolumn{2}{c}{$n=500$} \\
 \cmidrule(l{3pt}r{3pt}){3-4} \cmidrule(l{3pt}r{3pt}){5-6} \cmidrule(l{3pt}r{3pt}){7-8}
 &  & sLMg & sLM & sLMg & sLM & sLMg & sLM\\
 \cmidrule(l{3pt}r{3pt}){3-4} \cmidrule(l{3pt}r{3pt}){5-6} \cmidrule(l{3pt}r{3pt}){7-8}
-0.9 & -0.8 & 5.6 & 17.8 & 4.7 & 12.1 & 5.3 & 9.7\\
-0.6 & -0.8 & 8.6 & 16.7 & 6.6 & 13.4 & 5.1 & 10.4\\
-0.3 & -0.8 & 9.4 & 16.2 & 6.7 & 12.6 & 4.6 & 11.1\\
0.0 & -0.8 & 12.4 & 15.2 & 7.3 & 12.4 & 4.4 & 11.1\\
0.3 & -0.8 & 13.5 & 14.2 & 7.5 & 12.1 & 4.2 & 10.7\\
0.6 & -0.8 & 13.4 & 12.6 & 10.3 & 13.2 & 4.5 & 10.3\\
0.9 & -0.8 & 6.2 & 30.3 & 4.9 & 31.0 & 2.6 & 26.0\\
\addlinespace
-0.9 & -0.4 & 4.1 & 9.0 & 4.4 & 7.9 & 3.8 & 7.9\\
-0.6 & -0.4 & 4.1 & 6.1 & 3.6 & 6.2 & 2.4 & 7.6\\
-0.3 & -0.4 & 4.1 & 6.0 & 3.6 & 6.6 & 2.6 & 7.5\\
0.0 & -0.4 & 5.6 & 7.6 & 3.7 & 6.5 & 2.4 & 7.8\\
0.3 & -0.4 & 9.6 & 10.5 & 6.1 & 8.7 & 3.0 & 8.1\\
0.6 & -0.4 & 4.7 & 16.5 & 3.4 & 11.5 & 1.9 & 8.3\\
0.9 & -0.4 & 3.6 & 40.4 & 3.2 & 26.7 & 2.5 & 13.9\\
\addlinespace
-0.9 & 0.0 & 3.7 & 11.4 & 3.5 & 8.5 & 3.0 & 8.1\\
-0.6 & 0.0 & 3.4 & 5.4 & 3.2 & 5.9 & 2.1 & 7.4\\
-0.3 & 0.0 & 4.8 & 6.2 & 3.7 & 6.3 & 1.9 & 6.7\\
0.0 & 0.0 & 8.9 & 12.0 & 6.8 & 10.7 & 3.9 & 9.7\\
0.3 & 0.0 & 3.3 & 8.7 & 2.9 & 6.7 & 1.8 & 6.9\\
0.6 & 0.0 & 3.1 & 9.9 & 3.0 & 7.6 & 2.3 & 7.8\\
0.9 & 0.0 & 3.1 & 30.8 & 3.3 & 20.1 & 2.9 & 11.5\\
\addlinespace
-0.9 & 0.4 & 3.8 & 15.0 & 3.5 & 10.7 & 2.8 & 8.3\\
-0.6 & 0.4 & 5.2 & 7.6 & 3.9 & 6.9 & 2.2 & 6.9\\
-0.3 & 0.4 & 6.5 & 10.7 & 5.0 & 8.7 & 2.3 & 7.6\\
0.0 & 0.4 & 3.8 & 6.2 & 3.1 & 5.6 & 2.0 & 7.1\\
0.3 & 0.4 & 3.2 & 6.0 & 3.1 & 6.4 & 2.1 & 6.8\\
0.6 & 0.4 & 3.5 & 8.8 & 3.1 & 7.0 & 2.8 & 7.5\\
0.9 & 0.4 & 3.3 & 27.4 & 3.8 & 18.6 & 3.6 & 12.4\\
\addlinespace
-0.9 & 0.8 & 7.6 & 15.3 & 5.6 & 11.1 & 3.2 & 8.8\\
-0.6 & 0.8 & 8.1 & 10.0 & 5.2 & 8.3 & 3.2 & 8.3\\
-0.3 & 0.8 & 6.5 & 8.5 & 4.4 & 7.4 & 2.9 & 7.4\\
0.0 & 0.8 & 5.8 & 9.3 & 4.4 & 7.6 & 3.1 & 7.5\\
0.3 & 0.8 & 5.2 & 8.8 & 4.3 & 8.1 & 3.5 & 7.8\\
0.6 & 0.8 & 5.1 & 10.8 & 4.3 & 9.2 & 4.3 & 9.2\\
0.9 & 0.8 & 6.4 & 25.0 & 5.8 & 19.3 & 6.4 & 14.2\\
 \cmidrule(l{3pt}r{3pt}){3-4} \cmidrule(l{3pt}r{3pt}){5-6} \cmidrule(l{3pt}r{3pt}){7-8}
\end{tabular}

  \caption{Empirical size (percent) at nominal level $\alpha=5\%$ for the ARMA(1,1)-GARCH(1,1)  process of Eq.~(\ref{ARMAGARCHsim}). Case A.}\label{tab:A}
\end{table}

\begin{table}[H]
\centering
\small
\begin{tabular}{rrrrrrrr}
\multicolumn{1}{c}{$\phi_1$} & \multicolumn{1}{c}{$\theta_1$} & \multicolumn{2}{c}{$n=100$} & \multicolumn{2}{c}{$n=200$} & \multicolumn{2}{c}{$n=500$} \\
 \cmidrule(l{3pt}r{3pt}){3-4} \cmidrule(l{3pt}r{3pt}){5-6} \cmidrule(l{3pt}r{3pt}){7-8}
 &  & sLMg & sLM & sLMg & sLM & sLMg & sLM\\
 \cmidrule(l{3pt}r{3pt}){3-4} \cmidrule(l{3pt}r{3pt}){5-6} \cmidrule(l{3pt}r{3pt}){7-8}
-0.9 & -0.8 & 5.1 & 21.0 & 4.7 & 15.8 & 5.4 & 12.7\\
-0.6 & -0.8 & 9.6 & 20.5 & 7.2 & 15.7 & 6.4 & 15.0\\
-0.3 & -0.8 & 11.4 & 20.0 & 7.9 & 16.5 & 7.3 & 15.6\\
0.0 & -0.8 & 11.9 & 18.8 & 8.1 & 14.9 & 7.2 & 14.1\\
0.3 & -0.8 & 13.2 & 16.3 & 8.6 & 14.1 & 6.9 & 12.7\\
0.6 & -0.8 & 12.8 & 15.8 & 10.8 & 13.9 & 6.9 & 12.2\\
0.9 & -0.8 & 6.9 & 28.9 & 5.8 & 29.8 & 5.1 & 25.5\\
\addlinespace
-0.9 & -0.4 & 4.7 & 13.1 & 5.0 & 13.2 & 4.6 & 13.4\\
-0.6 & -0.4 & 5.3 & 9.8 & 5.4 & 11.8 & 5.6 & 14.5\\
-0.3 & -0.4 & 5.2 & 9.4 & 5.5 & 11.4 & 5.9 & 13.8\\
0.0 & -0.4 & 6.0 & 9.7 & 5.2 & 10.6 & 5.6 & 13.0\\
0.3 & -0.4 & 7.7 & 12.2 & 7.7 & 11.9 & 6.7 & 12.5\\
0.6 & -0.4 & 5.0 & 17.7 & 5.1 & 14.0 & 5.2 & 12.2\\
0.9 & -0.4 & 5.0 & 41.3 & 4.8 & 28.9 & 5.0 & 18.2\\
\addlinespace
-0.9 & 0.0 & 4.7 & 15.5 & 4.7 & 13.3 & 4.2 & 14.4\\
-0.6 & 0.0 & 4.8 & 8.8 & 4.8 & 10.2 & 5.2 & 13.1\\
-0.3 & 0.0 & 5.4 & 9.3 & 5.1 & 10.4 & 5.1 & 12.6\\
0.0 & 0.0 & 8.0 & 13.0 & 7.6 & 13.1 & 7.1 & 13.4\\
0.3 & 0.0 & 4.4 & 11.1 & 4.4 & 10.1 & 5.1 & 11.9\\
0.6 & 0.0 & 3.9 & 12.6 & 4.8 & 11.4 & 5.1 & 12.3\\
0.9 & 0.0 & 4.7 & 33.8 & 4.5 & 24.8 & 5.1 & 18.5\\
\addlinespace
-0.9 & 0.4 & 5.1 & 17.7 & 5.1 & 14.2 & 4.5 & 13.0\\
-0.6 & 0.4 & 5.9 & 10.1 & 4.9 & 9.6 & 4.8 & 10.8\\
-0.3 & 0.4 & 6.6 & 12.6 & 6.2 & 11.6 & 5.8 & 11.1\\
0.0 & 0.4 & 4.2 & 8.3 & 4.0 & 8.5 & 4.7 & 10.8\\
0.3 & 0.4 & 4.1 & 8.8 & 4.5 & 9.4 & 5.4 & 11.8\\
0.6 & 0.4 & 4.3 & 11.8 & 5.2 & 12.0 & 5.8 & 14.0\\
0.9 & 0.4 & 4.5 & 30.4 & 5.2 & 23.8 & 5.6 & 19.4\\
\addlinespace
-0.9 & 0.8 & 6.7 & 15.0 & 6.3 & 12.7 & 5.2 & 9.9\\
-0.6 & 0.8 & 7.6 & 11.1 & 6.7 & 10.0 & 5.2 & 8.6\\
-0.3 & 0.8 & 6.7 & 10.9 & 5.3 & 9.3 & 4.6 & 8.3\\
0.0 & 0.8 & 6.2 & 10.4 & 5.3 & 9.0 & 5.2 & 9.6\\
0.3 & 0.8 & 5.5 & 11.3 & 5.6 & 10.4 & 5.8 & 11.0\\
0.6 & 0.8 & 6.3 & 13.8 & 6.0 & 12.0 & 6.1 & 13.9\\
0.9 & 0.8 & 7.1 & 27.0 & 7.1 & 22.7 & 7.8 & 19.3\\
 \cmidrule(l{3pt}r{3pt}){3-4} \cmidrule(l{3pt}r{3pt}){5-6} \cmidrule(l{3pt}r{3pt}){7-8}
\end{tabular}
  \caption{Empirical size (percent) at nominal level $\alpha=5\%$ for the ARMA(1,1)-GARCH(1,1)  process of Eq.~(\ref{ARMAGARCHsim}). Case B.}\label{tab:B}
\end{table}

\begin{table}[H]
\centering
\small
\begin{tabular}{rrrrrrrr}
\multicolumn{1}{c}{$\phi_1$} & \multicolumn{1}{c}{$\theta_1$} & \multicolumn{2}{c}{$n=100$} & \multicolumn{2}{c}{$n=200$} & \multicolumn{2}{c}{$n=500$} \\
 \cmidrule(l{3pt}r{3pt}){3-4} \cmidrule(l{3pt}r{3pt}){5-6} \cmidrule(l{3pt}r{3pt}){7-8}
 &  & sLMg & sLM & sLMg & sLM & sLMg & sLM\\
 \cmidrule(l{3pt}r{3pt}){3-4} \cmidrule(l{3pt}r{3pt}){5-6} \cmidrule(l{3pt}r{3pt}){7-8}
-0.9 & -0.8 & 4.5 & 25.4 & 4.0 & 23.3 & 5.0 & 26.0\\
-0.6 & -0.8 & 6.6 & 25.8 & 4.7 & 27.4 & 5.1 & 32.7\\
-0.3 & -0.8 & 7.8 & 27.9 & 5.4 & 29.3 & 5.1 & 35.2\\
0.0 & -0.8 & 9.1 & 26.1 & 4.9 & 28.1 & 4.4 & 33.1\\
0.3 & -0.8 & 9.1 & 23.9 & 4.9 & 25.8 & 4.1 & 29.8\\
0.6 & -0.8 & 8.6 & 20.2 & 6.4 & 24.2 & 4.1 & 28.1\\
0.9 & -0.8 & 5.2 & 30.4 & 3.9 & 33.6 & 3.6 & 33.8\\
\addlinespace
-0.9 & -0.4 & 3.6 & 18.7 & 3.5 & 22.4 & 4.4 & 28.6\\
-0.6 & -0.4 & 3.8 & 17.3 & 3.4 & 23.1 & 3.3 & 32.6\\
-0.3 & -0.4 & 3.9 & 17.5 & 3.0 & 23.9 & 2.7 & 32.9\\
0.0 & -0.4 & 4.6 & 17.7 & 3.2 & 22.2 & 2.4 & 31.3\\
0.3 & -0.4 & 6.4 & 18.9 & 4.4 & 22.8 & 3.2 & 29.7\\
0.6 & -0.4 & 3.6 & 21.6 & 2.6 & 22.2 & 2.2 & 27.1\\
0.9 & -0.4 & 3.4 & 43.1 & 3.3 & 33.9 & 3.1 & 29.6\\
\addlinespace
-0.9 & 0.0 & 3.0 & 20.9 & 3.3 & 22.7 & 3.5 & 29.3\\
-0.6 & 0.0 & 3.1 & 15.8 & 2.7 & 21.3 & 2.6 & 31.4\\
-0.3 & 0.0 & 3.7 & 16.2 & 3.0 & 20.9 & 2.5 & 29.8\\
0.0 & 0.0 & 5.5 & 18.8 & 5.0 & 22.5 & 3.7 & 31.0\\
0.3 & 0.0 & 3.1 & 16.1 & 2.0 & 20.0 & 2.0 & 28.4\\
0.6 & 0.0 & 2.7 & 16.9 & 2.6 & 20.9 & 2.5 & 29.1\\
0.9 & 0.0 & 3.5 & 36.1 & 3.3 & 29.9 & 4.1 & 30.3\\
\addlinespace
-0.9 & 0.4 & 3.2 & 22.9 & 3.1 & 24.0 & 3.1 & 28.1\\
-0.6 & 0.4 & 4.2 & 16.6 & 3.3 & 20.2 & 2.9 & 28.0\\
-0.3 & 0.4 & 4.7 & 17.5 & 3.8 & 21.1 & 2.9 & 27.9\\
0.0 & 0.4 & 2.9 & 13.9 & 2.5 & 18.4 & 2.3 & 27.0\\
0.3 & 0.4 & 3.2 & 13.9 & 2.8 & 19.0 & 2.7 & 28.1\\
0.6 & 0.4 & 3.6 & 15.2 & 3.3 & 20.8 & 3.0 & 28.2\\
0.9 & 0.4 & 3.8 & 32.9 & 4.2 & 30.0 & 5.2 & 31.3\\
\addlinespace
-0.9 & 0.8 & 5.3 & 20.7 & 4.1 & 21.6 & 3.6 & 23.5\\
-0.6 & 0.8 & 5.0 & 16.9 & 3.9 & 19.2 & 3.6 & 22.3\\
-0.3 & 0.8 & 4.6 & 16.0 & 3.8 & 18.0 & 3.4 & 21.9\\
0.0 & 0.8 & 4.6 & 16.0 & 3.9 & 17.9 & 3.5 & 23.9\\
0.3 & 0.8 & 4.6 & 16.9 & 4.2 & 19.0 & 4.5 & 26.0\\
0.6 & 0.8 & 5.5 & 18.7 & 4.9 & 20.0 & 5.3 & 27.2\\
0.9 & 0.8 & 6.7 & 31.8 & 7.3 & 30.9 & 8.8 & 32.9\\
 \cmidrule(l{3pt}r{3pt}){3-4} \cmidrule(l{3pt}r{3pt}){5-6} \cmidrule(l{3pt}r{3pt}){7-8}
\end{tabular}

  \caption{Empirical size (percent) at nominal level $\alpha=5\%$ for the ARMA(1,1)-GARCH(1,1)  process of Eq.~(\ref{ARMAGARCHsim}). Case C.}\label{tab:C}
\end{table}

\subsection{Power of the tests}

\begin{table}[H]
\centering
\caption{Size corrected power (percent) of the sLMg and sLM tests, at nominal level $\alpha=5\%$ for the TARMA(1,1)-GARCH(1,1) process of Eq.~(\ref{TARMAGARCHsim}).}\label{tab:4}
\begin{tabular}{lrrrrrrr}
\multicolumn{1}{c}{} & \multicolumn{1}{c}{$\Psi$ } & \multicolumn{2}{c}{$n=100$} & \multicolumn{2}{c}{$n=200$} & \multicolumn{2}{c}{$n=500$} \\
\cmidrule(l{3pt}r{3pt}){3-4} \cmidrule(l{3pt}r{3pt}){5-6} \cmidrule(l{3pt}r{3pt}){7-8}
  &  & sLMg & sLM & sLMg & sLM & sLMg & sLM\\
\cmidrule(l{3pt}r{3pt}){3-4} \cmidrule(l{3pt}r{3pt}){5-6} \cmidrule(l{3pt}r{3pt}){7-8}
 & 0.0 & 5.0 & 5.0 & 5.0 & 5.0 & 5.0 & 5.0\\
 & -0.1 & 9.2 & 9.9 & 16.7 & 17.1 & 45.0 & 42.7\\
 & -0.3 & 25.9 & 27.2 & 53.8 & 55.3 & 96.7 & 95.5\\
 & -0.4 & 50.1 & 53.1 & 86.8 & 88.4 & 100.0 & 100.0\\
A & -0.6 & 71.9 & 77.0 & 97.8 & 98.7 & 100.0 & 100.0\\
 & -0.8 & 85.4 & 91.5 & 99.4 & 99.8 & 100.0 & 100.0\\
 & -0.9 & 90.4 & 96.5 & 99.3 & 99.9 & 100.0 & 100.0\\
 & -1.0 & 90.8 & 98.1 & 99.0 & 99.9 & 100.0 & 100.0\\
\addlinespace
 & 0.0 & 5.0 & 5.0 & 5.0 & 5.0 & 5.0 & 5.0\\
 & -0.1 & 8.4 & 7.3 & 11.5 & 11.6 & 27.2 & 25.1\\
 & -0.3 & 19.4 & 17.9 & 39.8 & 41.2 & 83.2 & 83.6\\
 & -0.4 & 44.9 & 44.2 & 76.9 & 81.2 & 99.5 & 99.7\\
B & -0.6 & 68.5 & 72.3 & 94.1 & 97.3 & 99.9 & 100.0\\
 & -0.8 & 83.2 & 89.8 & 94.7 & 99.1 & 98.8 & 100.0\\
 & -0.9 & 88.1 & 95.9 & 94.1 & 99.5 & 97.2 & 99.9\\
 & -1.0 & 91.1 & 97.7 & 94.4 & 99.6 & 98.0 & 99.9\\
\addlinespace
 & 0.0 & 5.0 & 5.0 & 5.0 & 5.0 & 5.0 & 5.0\\
 & -0.1 & 8.2 & 8.0 & 13.2 & 10.6 & 32.0 & 16.7\\
 & -0.3 & 17.8 & 17.6 & 38.5 & 31.0 & 83.8 & 62.2\\
 & -0.4 & 35.5 & 35.8 & 70.8 & 65.3 & 99.3 & 95.9\\
C & -0.6 & 55.3 & 60.2 & 90.1 & 89.6 & 100.0 & 99.8\\
 & -0.8 & 71.7 & 79.8 & 95.5 & 97.4 & 99.9 & 100.0\\
 & -0.9 & 78.4 & 91.2 & 95.1 & 98.9 & 99.5 & 99.9\\
 & -1.0 & 81.7 & 95.4 & 94.0 & 99.1 & 99.0 & 100.0\\
\cmidrule(l{3pt}r{3pt}){3-4} \cmidrule(l{3pt}r{3pt}){5-6} \cmidrule(l{3pt}r{3pt}){7-8}
\end{tabular}
\end{table}

\begin{table}[H]
\centering
\caption{Raw power (percent) of the sLMg and sLM tests, at nominal level $\alpha=5\%$ for the TARMA(1,1)-GARCH(1,1)  process of Eq.~(\ref{TARMAGARCHsim}).}\label{tab:4u}
\begin{tabular}{lrrrrrrr}
\multicolumn{1}{c}{} & \multicolumn{1}{c}{$\Psi$ } & \multicolumn{2}{c}{$n=100$} & \multicolumn{2}{c}{$n=200$} & \multicolumn{2}{c}{$n=500$} \\
\cmidrule(l{3pt}r{3pt}){3-4} \cmidrule(l{3pt}r{3pt}){5-6} \cmidrule(l{3pt}r{3pt}){7-8}
  &  & sLMg & sLM & sLMg & sLM & sLMg & sLM\\
\cmidrule(l{3pt}r{3pt}){3-4} \cmidrule(l{3pt}r{3pt}){5-6} \cmidrule(l{3pt}r{3pt}){7-8}
 & 0.0 & 3.7 & 5.6 & 3.5 & 6.2 & 2.4 & 7.0\\
 & -0.1 & 6.9 & 10.8 & 12.5 & 19.6 & 34.2 & 48.2\\
 & -0.3 & 20.9 & 28.9 & 47.6 & 58.3 & 94.0 & 96.7\\
 & -0.4 & 43.6 & 54.8 & 83.4 & 89.8 & 100.0 & 100.0\\
A & -0.6 & 66.2 & 78.4 & 97.0 & 98.9 & 100.0 & 100.0\\
 & -0.8 & 81.9 & 92.2 & 99.2 & 99.9 & 100.0 & 100.0\\
 & -0.9 & 87.9 & 96.7 & 99.2 & 99.9 & 100.0 & 100.0\\
 & -1.0 & 89.2 & 98.3 & 98.7 & 99.9 & 100.0 & 100.0\\
\addlinespace
 & 0.0 & 4.3 & 9.3 & 5.3 & 10.5 & 5.7 & 13.4\\
 & -0.1 & 7.4 & 13.0 & 12.0 & 20.1 & 29.1 & 43.0\\
 & -0.3 & 17.8 & 27.6 & 40.7 & 54.3 & 84.3 & 92.5\\
 & -0.4 & 42.6 & 56.3 & 77.5 & 88.7 & 99.6 & 99.9\\
B & -0.6 & 66.5 & 81.3 & 94.3 & 98.5 & 99.9 & 100.0\\
 & -0.8 & 82.1 & 94.1 & 94.8 & 99.5 & 98.9 & 100.0\\
 & -0.9 & 87.1 & 97.5 & 94.2 & 99.8 & 97.4 & 100.0\\
 & -1.0 & 90.5 & 98.6 & 94.6 & 99.8 & 98.2 & 100.0\\
\addlinespace
 & 0.0 & 3.4 & 13.8 & 3.1 & 19.0 & 3.1 & 28.4\\
 & -0.1 & 5.9 & 19.0 & 9.3 & 31.2 & 25.9 & 56.8\\
 & -0.3 & 13.5 & 34.2 & 31.5 & 61.2 & 79.5 & 92.8\\
 & -0.4 & 29.6 & 57.1 & 63.4 & 87.6 & 98.9 & 99.7\\
C & -0.6 & 49.4 & 79.2 & 86.8 & 97.7 & 100.0 & 100.0\\
 & -0.8 & 66.3 & 91.3 & 94.0 & 99.4 & 99.9 & 100.0\\
 & -0.9 & 74.3 & 96.9 & 93.6 & 99.8 & 99.4 & 100.0\\
 & -1.0 & 78.0 & 98.5 & 92.5 & 99.8 & 98.7 & 100.0\\
\cmidrule(l{3pt}r{3pt}){3-4} \cmidrule(l{3pt}r{3pt}){5-6} \cmidrule(l{3pt}r{3pt}){7-8}
\end{tabular}
\end{table}

\subsection{Power of the sLMg test in presence of measurement error}\label{Supp:merr}

We study the power of the sLMg test in presence of measurement error by simulating from the following TAR$(1)$-GARCH$(1,1)$ model:
\begin{align}\label{TARGARCHsim}
 X_t &= 0.5 + 0.5 X_{t-1} + \left(\varphi_0 + \varphi_1 X_{t-1} + \right)I(X_{t-1}\leq 0) + \eps_t.\nonumber\\
 \eps_t&=\sqrt{h_t}z_t \quad
h_t=a_0 + a_1\eps^2_{t-1} + b_1 h_{t-1}.
\end{align}
where $\varphi_{0}=\varphi_1 = 0.5 + \Psi$, where $$\Psi = (0.00, -0.15, -0.30, -0.45, -0.60, -0.75, -0.90, -1.05).$$
As above, we combine these with the following parameters for the GARCH specification:  $(a_0,a_1,b_1) = (1,0.1,0.8)$ (case A), $(1,0.4,0.4)$ (case B), $(1,0.8,0.1)$ (case C). We add measurement noise as above: $Y_t = X_t + \eta_t$, where $\eta_t$ and the SNR are as above. The size corrected power of the sLMg test is reported in Table~\ref{tab:5}. As expected, the presence of high levels of measurement noise impinge negatively upon the power of the test so that larger sample sizes are required to compensate for it.

\begin{table}
\centering
\caption{Size corrected power (percent) of the sLMg test, at nominal level $\alpha=5\%$ for the TAR(1)-GARCH(1,1)  process of Eq.~(\ref{TARGARCHsim}) with measurement error.}\label{tab:5}
\begin{tabular}{lrrrrrrrrrrrrr}
\multicolumn{1}{c}{} & \multicolumn{1}{c}{$\Psi$} & \multicolumn{4}{c}{$n=100$} & \multicolumn{4}{c}{$n=200$} & \multicolumn{4}{c}{$n=500$} \\
 \cmidrule(l{3pt}r{3pt}){3-6} \cmidrule(l{3pt}r{3pt}){7-10} \cmidrule(l{3pt}r{3pt}){11-14}
  &  & $\infty$ & 50 & 10 & 5 & $\infty$ & 50 & 10 & 5 & $\infty$ & 50 & 10 & 5\\
 \cmidrule(l{3pt}r{3pt}){3-6} \cmidrule(l{3pt}r{3pt}){7-10} \cmidrule(l{3pt}r{3pt}){11-14}
 & 0.0 & 5.0 & 4.9 & 4.9 & 4.9 & 4.9 & 4.9 & 4.9 & 5.0 & 5.0 & 4.9 & 5.0 & 5.0\\
 & -0.1 & 6.8 & 6.9 & 4.4 & 4.7 & 7.0 & 6.7 & 5.8 & 6.0 & 10.3 & 9.5 & 9.4 & 7.4\\
 & -0.3 & 8.5 & 10.1 & 8.2 & 5.4 & 15.4 & 15.9 & 9.9 & 8.8 & 35.7 & 32.2 & 27.3 & 15.7\\
 & -0.4 & 15.3 & 14.8 & 10.7 & 8.2 & 25.4 & 24.6 & 18.9 & 11.5 & 66.8 & 60.5 & 46.9 & 26.2\\
A & -0.6 & 19.4 & 19.6 & 15.4 & 10.0 & 38.9 & 36.5 & 25.7 & 15.6 & 89.8 & 85.5 & 74.2 & 41.4\\
 & -0.8 & 28.8 & 29.9 & 21.1 & 11.0 & 58.7 & 52.3 & 38.8 & 19.4 & 97.1 & 95.7 & 88.2 & 60.3\\
 & -0.9 & 40.4 & 39.6 & 28.1 & 15.3 & 70.9 & 67.1 & 52.4 & 30.1 & 98.7 & 98.5 & 94.6 & 71.8\\
 & -1.0 & 45.4 & 46.4 & 33.4 & 17.1 & 79.9 & 75.7 & 59.3 & 36.1 & 98.4 & 97.2 & 93.7 & 77.2\\
\addlinespace
 & 0.0 & 4.9 & 4.9 & 4.9 & 4.9 & 5.0 & 5.0 & 5.0 & 5.0 & 4.9 & 5.0 & 5.0 & 5.0\\
 & -0.1 & 6.0 & 6.3 & 5.3 & 5.6 & 5.9 & 5.4 & 6.1 & 6.4 & 6.6 & 8.7 & 7.3 & 6.0\\
 & -0.3 & 9.7 & 8.2 & 7.7 & 6.7 & 10.1 & 10.1 & 9.6 & 8.1 & 21.8 & 23.8 & 15.0 & 9.8\\
 & -0.4 & 15.9 & 15.9 & 13.1 & 9.4 & 24.3 & 21.8 & 15.6 & 10.8 & 52.4 & 50.6 & 34.2 & 21.7\\
B & -0.6 & 29.5 & 24.9 & 17.8 & 12.7 & 45.3 & 42.0 & 30.9 & 18.8 & 72.4 & 71.4 & 53.1 & 33.2\\
 & -0.8 & 45.0 & 39.5 & 25.1 & 16.8 & 67.9 & 59.9 & 43.7 & 25.4 & 87.5 & 86.5 & 77.5 & 54.4\\
 & -0.9 & 66.2 & 60.1 & 41.0 & 23.5 & 86.6 & 79.2 & 65.9 & 41.0 & 98.0 & 97.3 & 92.4 & 75.7\\
 & -1.0 & 79.8 & 73.8 & 52.5 & 29.3 & 94.5 & 93.9 & 80.0 & 52.3 & 99.9 & 99.7 & 98.3 & 89.6\\
\addlinespace
 & 0.0 & 5.0 & 4.9 & 4.9 & 4.9 & 5.0 & 5.0 & 5.0 & 5.0 & 5.0 & 5.0 & 5.0 & 5.0\\
 & -0.1 & 5.4 & 5.7 & 5.1 & 6.4 & 7.7 & 5.2 & 5.9 & 5.7 & 6.9 & 6.5 & 5.7 & 4.9\\
 & -0.3 & 7.6 & 7.7 & 8.5 & 8.0 & 12.7 & 10.1 & 9.2 & 7.8 & 21.5 & 18.9 & 13.4 & 11.1\\
 & -0.4 & 12.7 & 12.1 & 13.6 & 9.8 & 22.2 & 18.4 & 15.2 & 10.0 & 52.1 & 47.5 & 32.6 & 21.6\\
C & -0.6 & 22.6 & 21.4 & 16.8 & 12.0 & 37.8 & 30.7 & 22.8 & 14.8 & 71.6 & 68.4 & 48.1 & 28.4\\
 & -0.8 & 31.6 & 28.8 & 22.3 & 14.8 & 55.5 & 46.1 & 33.7 & 22.5 & 82.9 & 78.9 & 59.1 & 38.9\\
 & -0.9 & 42.6 & 38.6 & 29.3 & 19.7 & 67.2 & 57.8 & 40.6 & 27.8 & 85.7 & 84.0 & 68.7 & 46.6\\
 & -1.0 & 50.8 & 47.2 & 36.2 & 19.6 & 77.8 & 69.1 & 51.7 & 32.8 & 92.3 & 89.8 & 79.4 & 54.2\\
 \cmidrule(l{3pt}r{3pt}){3-6} \cmidrule(l{3pt}r{3pt}){7-10} \cmidrule(l{3pt}r{3pt}){11-14}
\end{tabular}
\end{table}

\renewcommand\thefigure{\thesection.\arabic{figure}}
\setcounter{figure}{0}
\renewcommand\thetable{\thesection.\arabic{table}}
\setcounter{table}{0}
\section{Analysis of Italian strikes}\label{SMsec:strikes}

\subsection{Additional plots}\label{Supp:Graph}

\begin{figure}[H]
\centering
\includegraphics[width=0.9\linewidth,keepaspectratio]{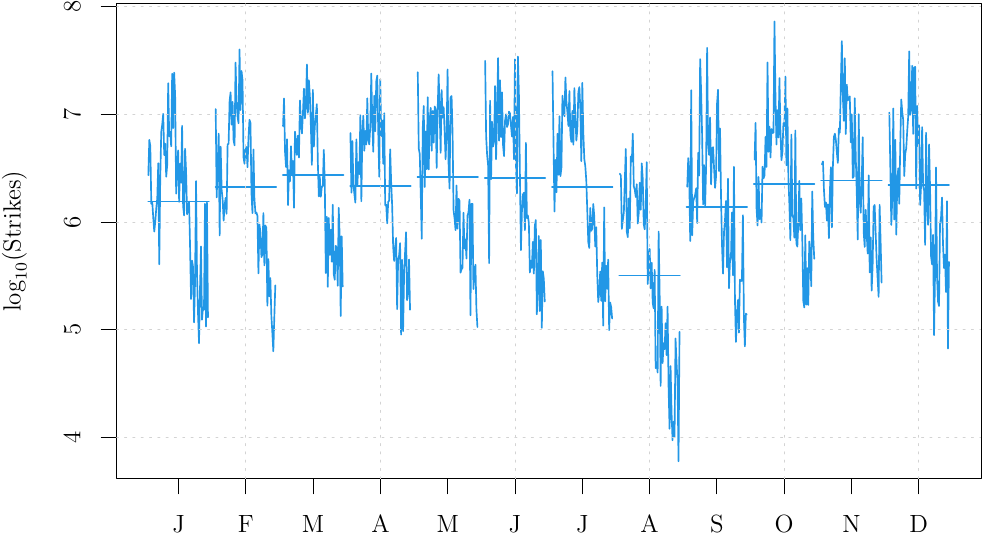}
\caption{Monthplot of the Italian Strikes (in $log_{10}$) from January 1949 to December 2009.}\label{fig:2}
\end{figure}
\begin{figure}[H]
\centering
\includegraphics[width=0.45\linewidth,keepaspectratio]{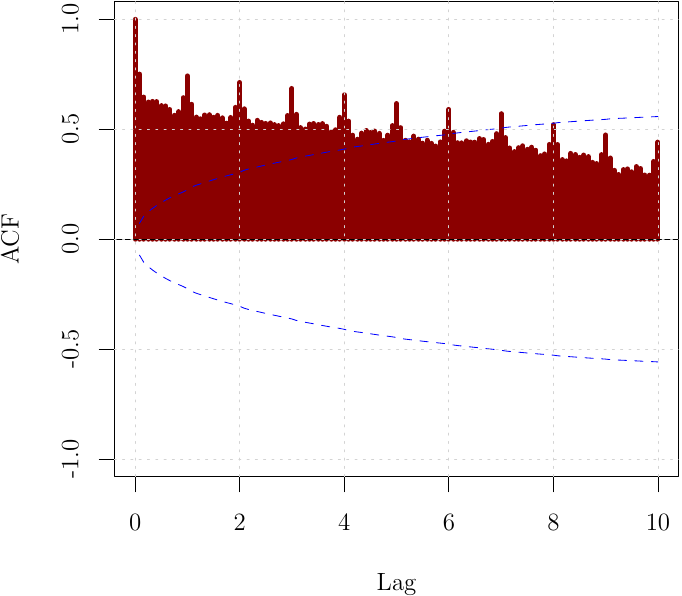}
\includegraphics[width=0.45\linewidth,keepaspectratio]{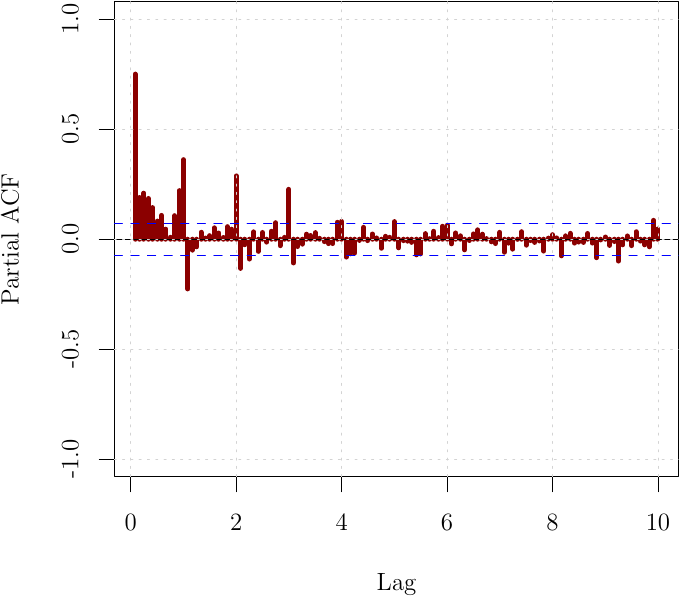}
	\caption{Correlograms of the Italian Strikes (in $log_{10}$) from January 1949 to December 2009. Autocorrelation function, with rejection bands under the hypothesis of a MA process (left). Partial autocorrelation function (right).}\label{fig:3}
\end{figure}
\subsection{ARIMA-GARCH modelling}\label{Supp:GARCH}
\setcounter{equation}{0}
\renewcommand\theequation{SM.\arabic{equation}}

In this Section we use a linear ARIMA specification to model the conditional mean and a GARCH(1,2) model for the conditional variance. The model includes two dummy variables: one for September 1953 ($D_1$), and the second for September-November 1969 ($D_2$).  In these two dates, strong increases in the strikes variable are observed, due to particularly intense strike activity, stimulated by the \emph{conglobamento} (the restructuring of wage scales) in 1953 and by the \emph{autunno caldo}, literally translated as the ``hot autumn'' in 1969, one of the largest mass movement in labour history for many countries \citep{franzosibook}.

\begin{figure}[H]
\centering
\includegraphics[width=0.8\linewidth,keepaspectratio]{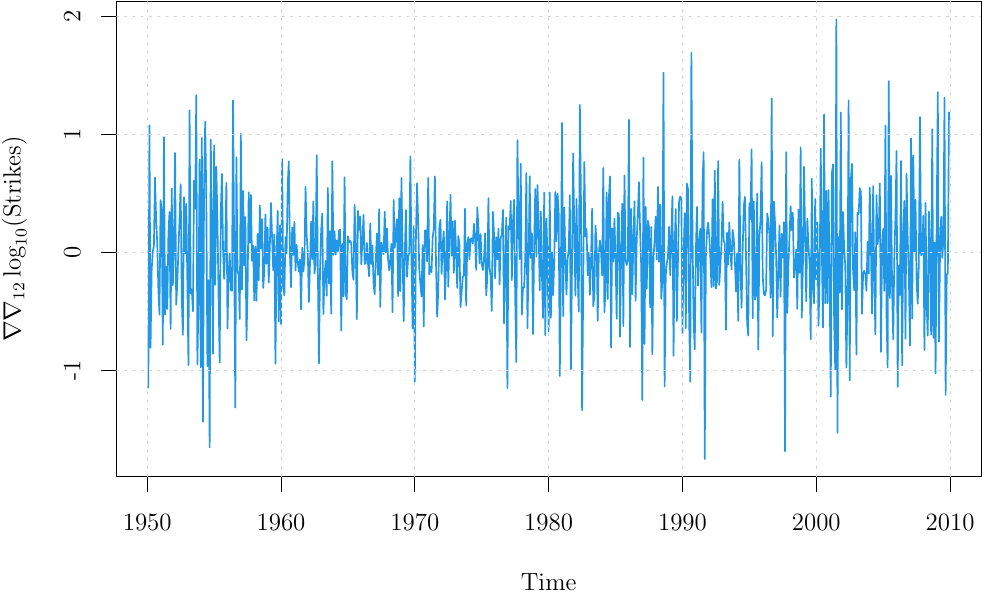}
\caption{Time series of the seasonal differences of the Italian Strikes (in $log_{10}$) from January 1949 to December 2009.}\label{fig:6}
\end{figure}

\begin{figure}[H]
\centering
\includegraphics[width=0.45\linewidth,keepaspectratio]{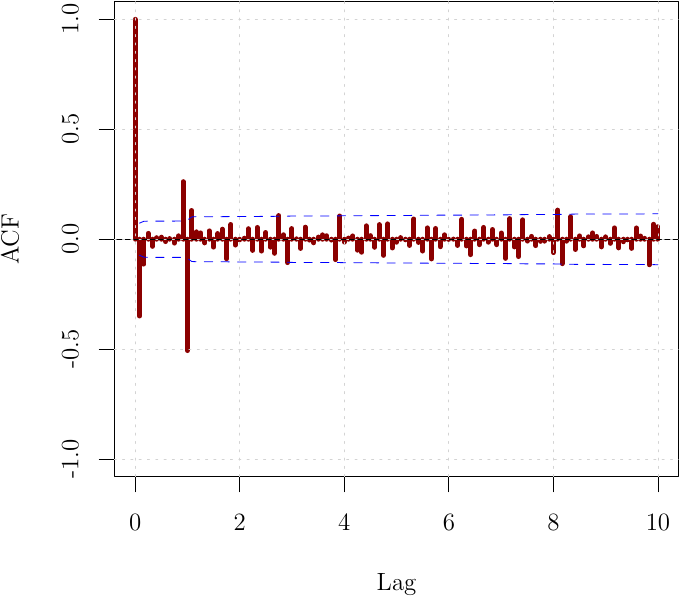}
\includegraphics[width=0.45\linewidth,keepaspectratio]{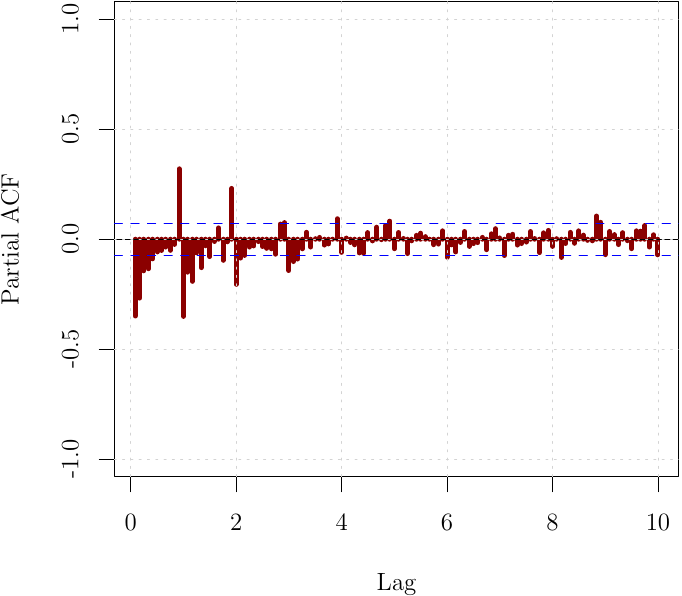}
\caption{Correlograms of the differenced series $\nabla\nabla_{12}\log_{10}$(Strikes). Autocorrelation function, with rejection bands under the hypothesis of a MA process (left). Partial autocorrelation function (right).}\label{fig:7}
\end{figure}

\begin{figure}[H]
\centering
\includegraphics[width=0.45\linewidth,keepaspectratio]{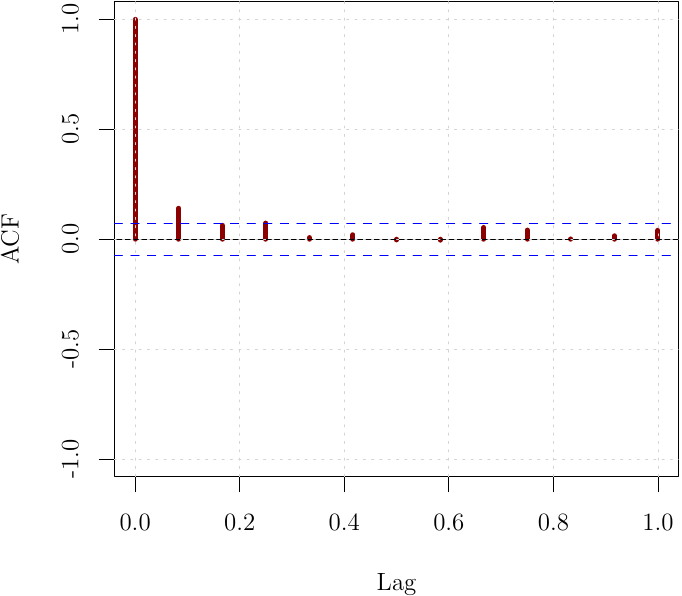}
\includegraphics[width=0.45\linewidth,keepaspectratio]{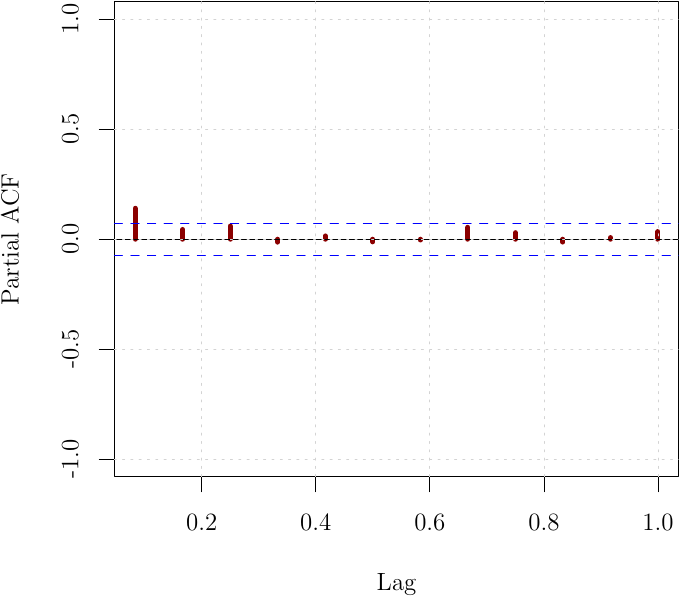}
\caption{Correlograms of the squared residuals of the ARIMA  model~(\ref{eq:arima}). Autocorrelation function, with rejection bands under the hypothesis of an MA process (left). Partial autocorrelation function (right).}\label{fig:8}
\end{figure}

\begin{figure}[H]
\centering
\includegraphics[width=0.45\linewidth,keepaspectratio]{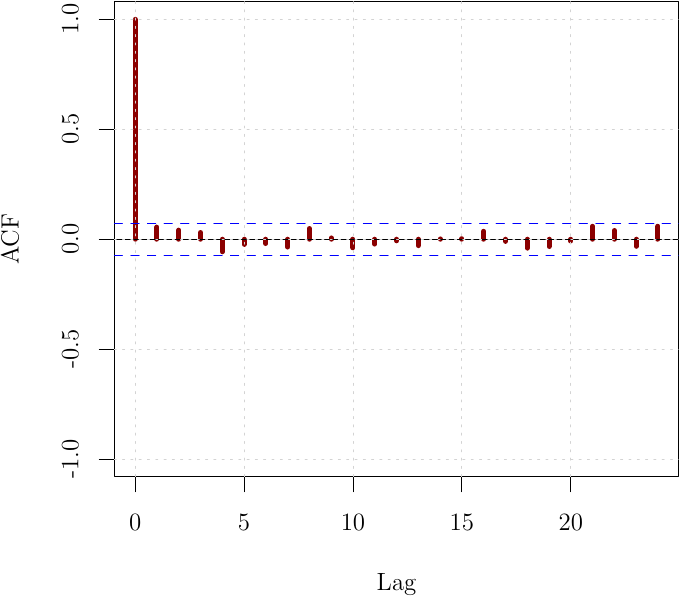}
\includegraphics[width=0.45\linewidth,keepaspectratio]{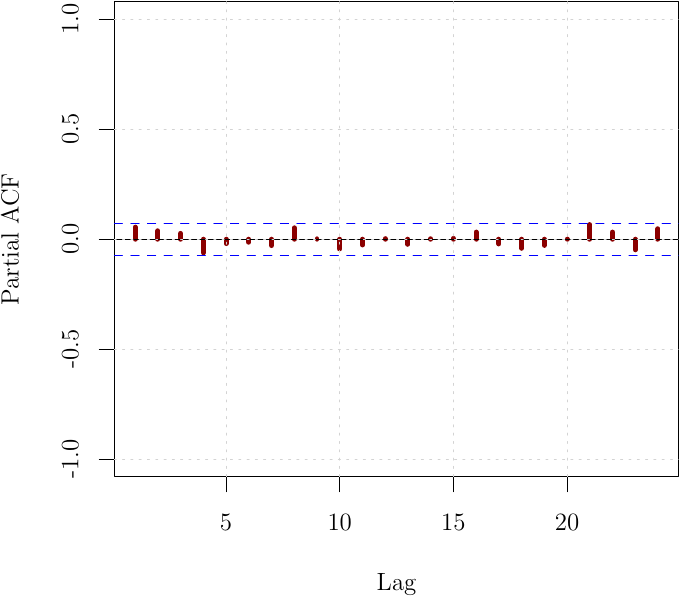}
\caption{Correlograms of the squared residuals of the ARIMA-GARCH model~(\ref{eq:arimag}). Autocorrelation function, with rejection bands under the hypothesis of an MA process (left). Partial autocorrelation function (right).}\label{fig:garch_res}
\end{figure}

If one adopts a linear specification then both a regular and a seasonal difference are needed. The series of the seasonal differences of the Italian Strikes (in $log_{10}$) is reported in Figure \ref{fig:6}, while Figure~\ref{fig:7} reports the correlograms of the same series.
The inspection of the correlograms points at the presence of both regular and seasonal MA components. Hence, we perform a two-stage fit over the following specification
\begin{align}
	(1-\phi_1 B)\nabla\nabla_{12} X_t &= (1+\theta_1 B+\theta_2 B^2)(1+\Theta_1 B)\eps_t + \delta_{1}D_1 + \delta_{2}D_2;\label{eq:arima}\\
	\eps_t &=   \sqrt{h_t}\,z_t; \quad
	h_t = \alpha_0 + \alpha_1 \eps^2_{t-1} + \beta_1 h_{t-1} + \beta_2 h_{t-2}\label{eq:arimag},
\end{align}
\noindent
where $\{z_t\}$ is a i.i.d. process, $B$ is the backshift operator, and $\nabla$ is the difference operator. We start by modelling the conditional mean by means of an ARIMA$(1,1,2)(0,1,1)_{12}$. The results are reported in Table~\ref{tab:arima}(left).
\begin{table}[ht]
	\centering
\caption{Model fit for the ARIMA$(1,1,2)(0,1,1)_{12}$--GARCH(1,2) model of Eqs.~(\ref{eq:arima}),(\ref{eq:arimag}).}\label{tab:arima}
	\begin{tabular}{rrr}
		 & Coefficients & s.e.  \\
		\cmidrule(lr){2-3}
		$\phi_1$ 	& 	0.62 	& 	0.10 \\
		$\theta_1$  &  -1.21    &   0.12 \\
		$\theta_2$ 	& 	0.25 	&	0.11 \\
		$\Theta_1$  &  -0.88	&  	0.02 \\
		$\delta_1$ 	& 	1.04	& 	0.29 \\
		$\delta_2$ 	& 	0.67	& 	0.22 \\
		\cmidrule(lr){2-3}
		Log likelihood  & -208.47 &  \\
		\cmidrule(lr){2-3}
		AIC             & 430.95  &  \\
		\cmidrule(lr){2-3}
		$\sigma^2$      & 0.10   &  \\
		\cmidrule(lr){2-3}
	\end{tabular}
	\begin{tabular}{rrr}
		 & Coefficients & s.e.  \\
		\cmidrule(lr){2-3}
		$\alpha_0$  &   0.001	&  	0.001 \\
		$\alpha_1$ 	& 	0.047	& 	0.020 \\
		$\beta_1$ 	& 	0.160	& 	0.043 \\
		$\beta_2$ 	& 	0.779	& 	0.042 \\
		\cmidrule(lr){2-3}
		Log likelihood  & -184.01 &  \\
		\cmidrule(lr){2-3}
	\end{tabular}
\end{table}
The Ljung-Box test applied to the squared residuals reveals the presence of an unaccounted ARCH effect ($X^2 = 51.543$, df = 24, $p$-value = 0.0009). This is also evident from the inspection of the correlograms of the squared residuals, shown in Figure \ref{fig:8}.
This prompts us to model the residuals with a GARCH(1,2) model. The results are shown in Table~\ref{tab:arima}(right). The correlograms of the squared standardized residuals, reported in Figure \ref{fig:garch_res}, show that no residual ARCH effect is present.

\begin{figure}
\centering
\includegraphics[width=0.45\linewidth,keepaspectratio]{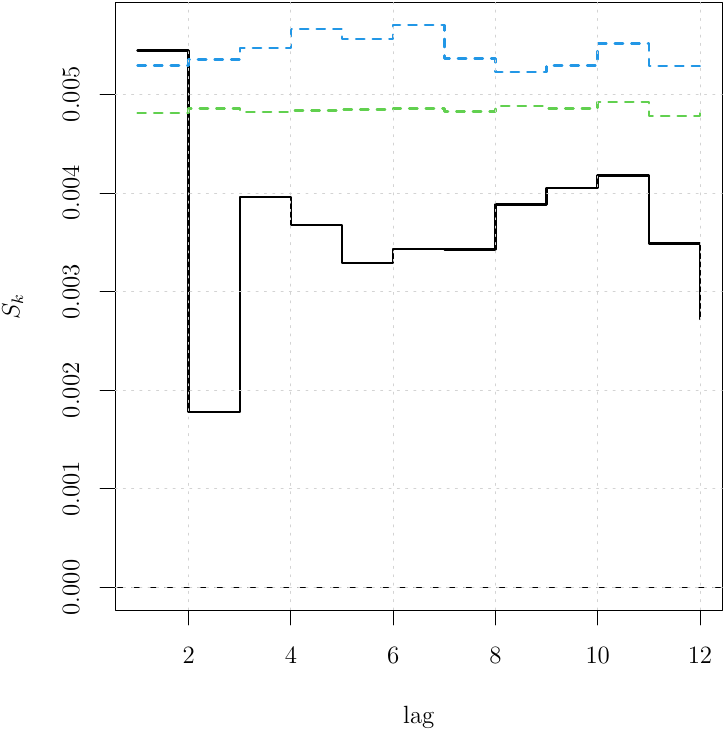}
\includegraphics[width=0.45\linewidth,keepaspectratio]{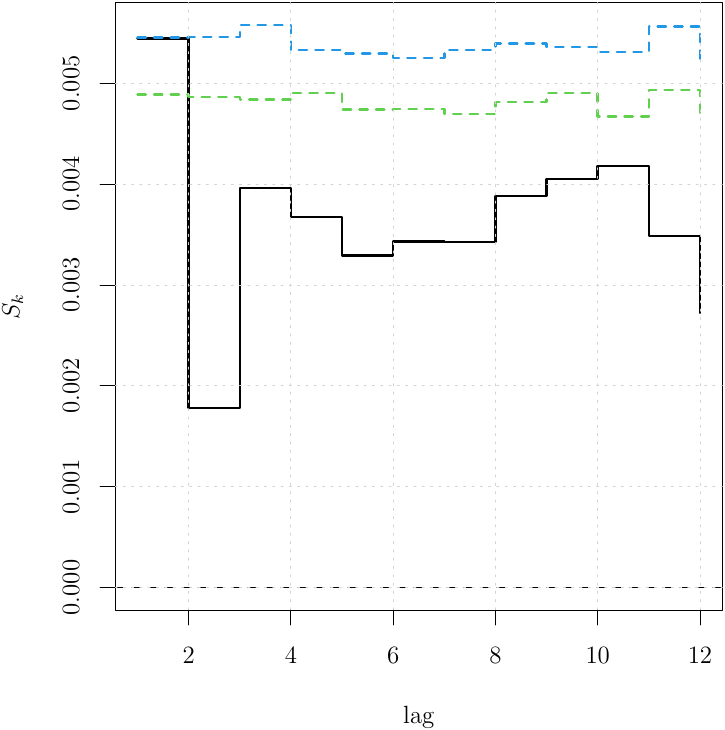}
			
\caption{Entropy-based metric up to lag 12, computed on the residuals of the ARIMA-GARCH model of Eq.~(\ref{eq:arimag}). The dashed lines correspond to bootstrap rejection bands at levels 95\% (green) and 99\% (blue) under the null hypothesis of independence (left panel) and linearity (right panel).}\label{fig:9}
\end{figure}

However, if we apply the entropy-based test for (nonlinear) serial dependence of \cite{Gia15} up to lag 12, see Figure~\ref{fig:9}, we observe a rejection at lag 1 at 1\% level both under the null of independence (left) and linearity (right). This points to an unaccounted dependence that requires a nonlinear modelling approach.

\subsection{Diagnostics for the TARMA-GARCH model}\label{Supp:Diag}

\begin{figure}[H]
\centering
\includegraphics[width=0.45\linewidth,keepaspectratio]{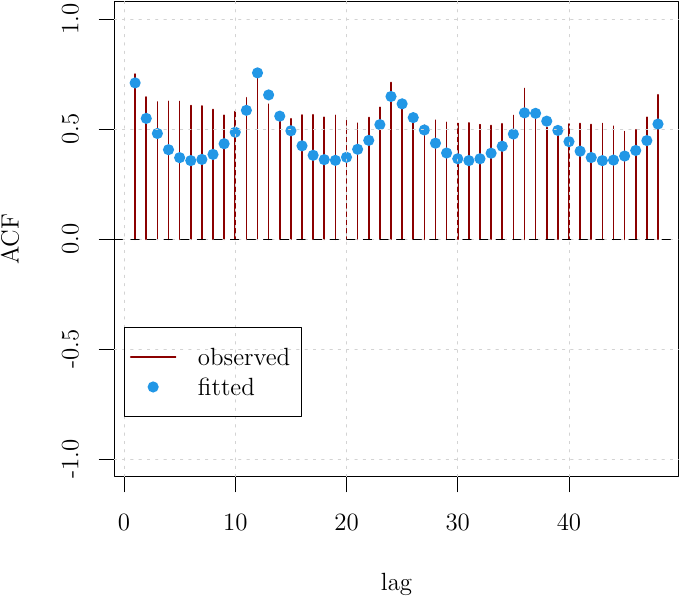}
\includegraphics[width=0.45\linewidth,keepaspectratio]{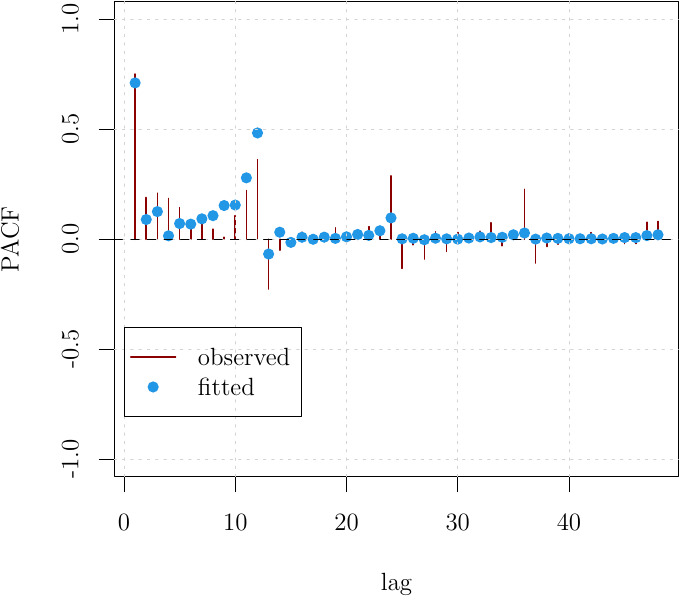}
\caption{Global (left) and partial (right) autocorrelation functions: empirical (red line), theoretical, simulated from the TARMA fit of Eq~(\ref{eq:TG}) (blue dots).}\label{fig:11}
\end{figure}

%
In the following, we provide evidence that the TARMA-GARCH model of Eq~(\ref{eq:TG}),  proposed in Section \ref{sec:real} is able to capture the main features of the strikes series. Even though in the nonlinear time series framework the autocorrelation function does not play the same central role as in the ARMA framework, a fitted nonlinear model should be able to reproduce the main correlation structure of the series. This is shown in Figure~\ref{fig:11}, where the correlograms derived from the simulated series are superimposed as blue dots onto the sample correlograms. The main structure is reproduced by the TARMA fit. The inclusion of additional terms in the model allows us to account for the slower decay of the global autocorrelation function and for the seasonality present in the partial correlogram at the expense of parsimony and interpretability.

\begin{figure}[H]
\centering
\includegraphics[width=0.45\linewidth,keepaspectratio]{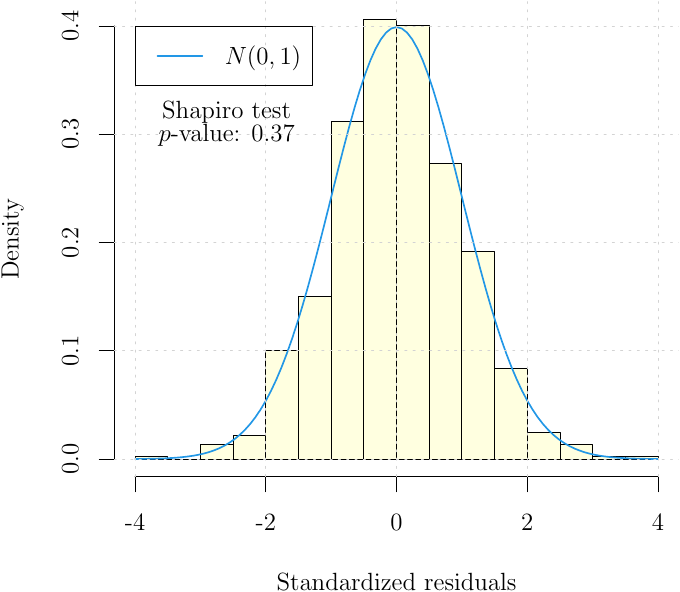}
\caption{Histogram of the standardized residuals from the TARMA fit of Eq~(\ref{eq:TG}). The smooth blue curve is the standard normal density and the $p$-value of the Shapiro-Wilk normality test is also reported.}\label{fig:12}
\end{figure}
\begin{figure}[H]
\centering
\includegraphics[width=0.45\linewidth,keepaspectratio]{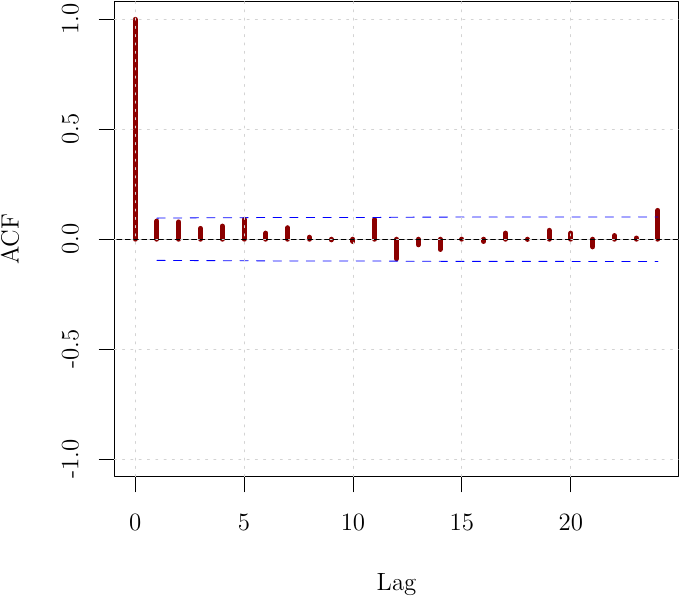}
\includegraphics[width=0.45\linewidth,keepaspectratio]{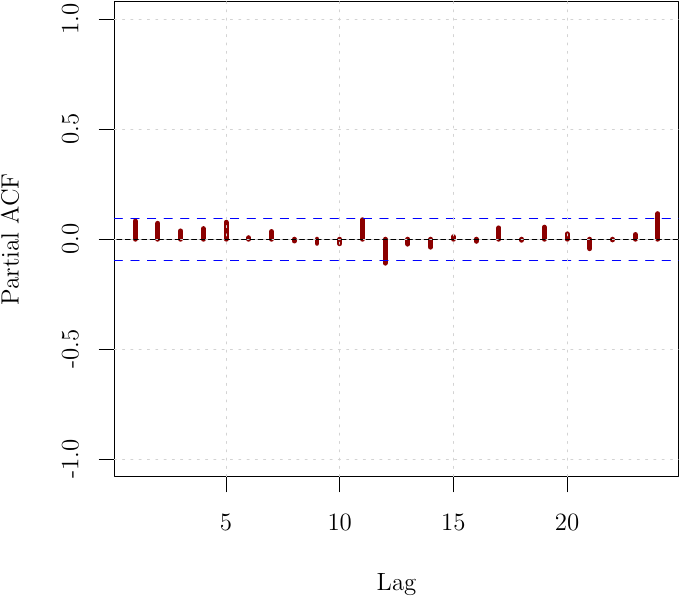}
\caption{Correlograms of the residuals of the TARMA fit of Eq~(\ref{eq:TG}). Autocorrelation function, with rejection bands under the hypothesis of a MA process (left). Partial autocorrelation function (right).}\label{fig:13}
\end{figure}
\begin{figure}[H]
\centering
\includegraphics[width=0.45\linewidth,keepaspectratio]{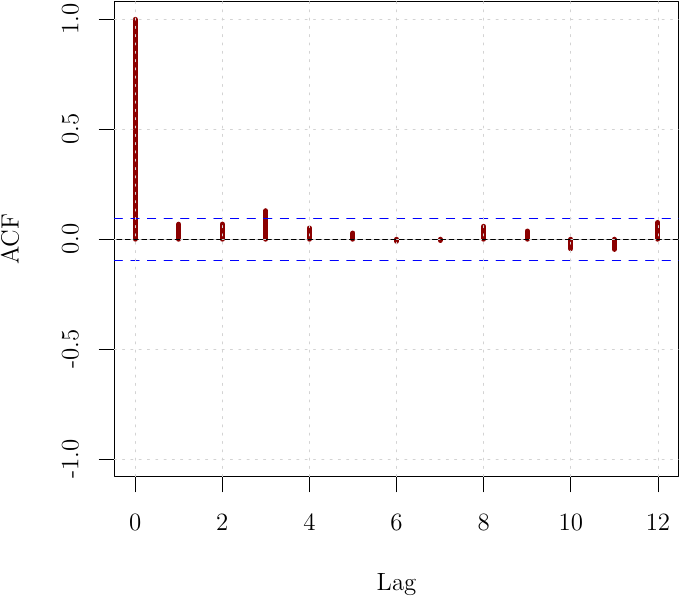}
\includegraphics[width=0.45\linewidth,keepaspectratio]{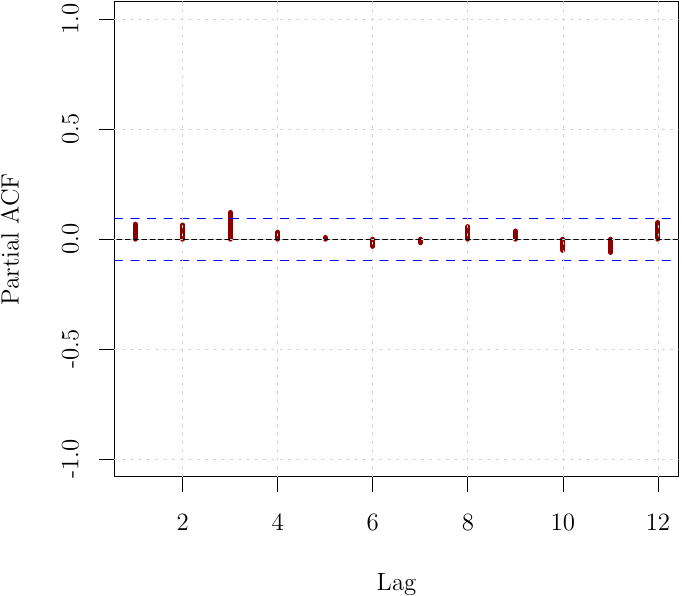}
\caption{Correlograms of the squared residuals of the TARMA model of Eq~(\ref{eq:TG}). Autocorrelation function (left). Partial autocorrelation function (right).}\label{fig:14}
\end{figure}

\noindent
The correlograms of the standardized residuals do not show significant deviations from the white noise hypothesis, see Figure~\ref{fig:13}.
\noindent
On the other hand, the correlograms of the squared standardized residuals (see Figure~\ref{fig:14}) and the Ljung-Box portmanteau test point at a weak but significant ARCH effect (X-squared = 32.24, df = 1, p-value = 1.36e-08)

We fit a GARCH(1,1) model upon the residuals of the series. The results are shown in Eq.~(\ref{garch.fit}). The diagnostic analysis does not show any unaccounted residual dependence, see Figure~\ref{fig:10}. We also test for the presence of cross-dependence between the residuals of the fitted TARMA-GARCH model and the covariates introduced previously, namely \emph{salary}: Monthly index of salaries of industrial workers and \emph{price}:  Monthly index of consumer prices. All the series have been prewhitened and the confidence bands under independence are obtained though random permutation. Details on the cross entropy measure can be found in \cite{Gia23b}.

\begin{figure}
\centering
\includegraphics[width=0.45\linewidth,keepaspectratio]{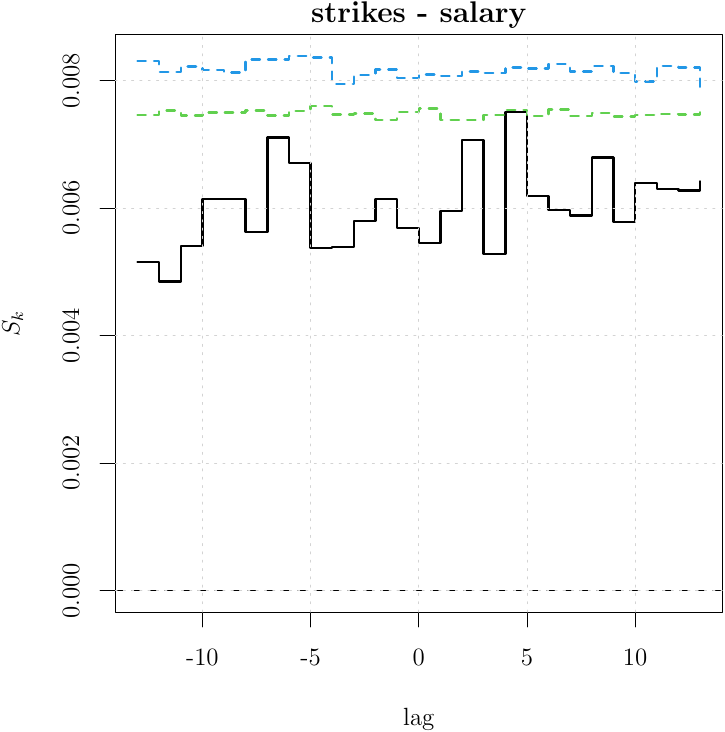}
\includegraphics[width=0.45\linewidth,keepaspectratio]{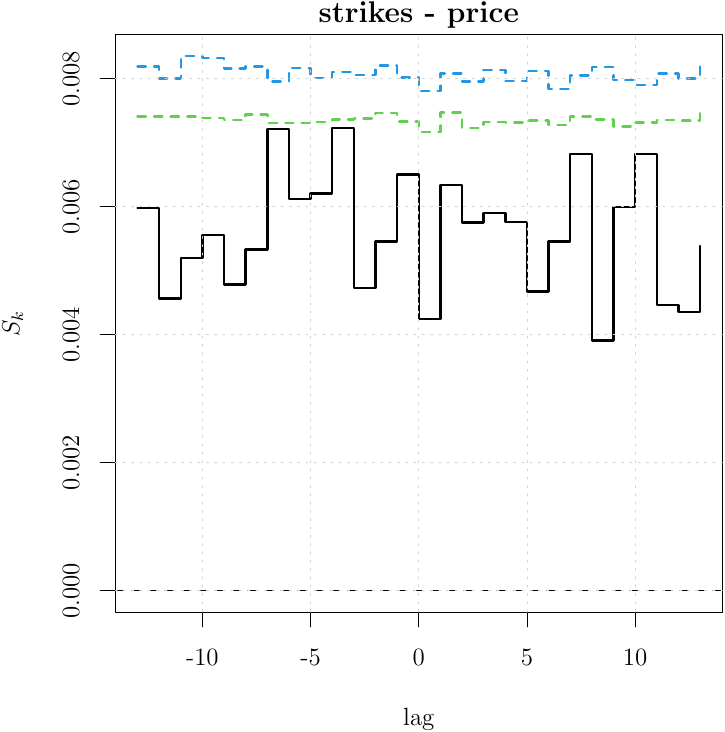}
			
\caption{Cross Entropy metric up to lag 12, between the residuals of the TARMA-GARCH model of Eqs.~(\ref{tarma.fit})--(\ref{garch.fit}) and \emph{salary} (left panel) and \emph{price} (right panel). The dashed lines correspond to permutation rejection bands at levels 95\% (green) and 99\% (blue) under the null hypothesis of cross-independence. The series have been prewhitened using a AR filter.}\label{fig:15}
\end{figure}


%
%

\end{document}